%% file: hydro-review.tex
\documentclass[final,3p]{elsarticle}

\usepackage{amssymb,custom}

\journal{Progress in Particle and Nuclear Physics}

\begin{document}

\begin{frontmatter}



\title{Recent developments in relativistic hydrodynamic fluctuations}

\author{G\"ok\c ce Ba\c sar}
\address{Department of Physics, University of North Carolina, Chapel Hill, NC, 27599, USA}


\begin{abstract}
The study of thermal fluctuations in relativistic hydrodynamics has led to numerous important developments in the last decade. We present a bird's eye view of the recent advances on the theory of fluctuations on three fronts; stochastic hydrodynamics, hydro-kinetics where fluctuations are included as additional modes that satisfy deterministic evolution equations, and effective field theory formulation of relativistic hydrodynamics. We compare and contrast these different but complimentary frameworks and highlight various recent progresses in each of them. 

\end{abstract}


\end{frontmatter}
\tableofcontents

\section{ Introduction }
 \input intro
 
 \section{ Hydrodynamic fluctuations in equilibrium }\label{sec:equilibrium}
 \input equilibrium

 \section{Stochastic hydrodynamics }\label{sec:stochastic}
\input stochastic

\section{Deterministic approach: hydro-kinetics}\label{sec:hydrokinetics}

\input hydrokinetic

\section{Effective field theory of fluctuating hydrodynamics}\label{sec:effective_action}
  \input effective

\section{Conclusions}\label{sec:conclusions}
  \input conclusions


\section*{Acknowledgements}
The author is supported by the National Science Foundation CAREER Award PHY-2143149.

\bibliographystyle{elsarticle-num} 
 \bibliography{hydro-review}


%
%
%
\end{document}

%% file: intro.tex
Hydrodynamics provides a universal description of macroscopic systems near thermal equilibrium in the long wavelength, small frequency limit. In this limit, regardless of what the underlying microscopic, short distance physics is, the dynamics of the system is governed by conservation laws. This is because throughout processes that happen in a macroscopic scale, the non-conserved quantities relax quickly towards their equilibrium values. In contrast, conserved quantities cannot be destroyed locally and they can only reach equilibrium through transport processes like diffusion which are much slower. 

In essence, hydrodynamics is a set of conservation equations that characterize how these slow modes evolve in time and eventually reach global equilibrium. Each conservation equation involves a conserved density and flux. The fluxes are expressed in terms of the conserved densities via the constitutive relations which close the system of conservation equations. In modern language, the constitutive relations are expressed as a derivative expansion where the leading term contains no derivatives (ideal hydrodynamics), the next term, one derivative (viscous hydrodynamics) and so on. In relativistic hydrodynamics \cite{Landau:2013fluid}, the subject of this review, the derivative expansion is constrained by the underlying Lorentz symmetry \cite{Romatschke:2017ejr}. The set of conservation equations constitutes system of partial differential equation which describe the evolution of the fluid in time. 

This formulation of hydrodynamics leaves out the role of \textit{thermal fluctuations}. As any macroscopic system, a fluid exhibits thermal fluctuations. In non-relativistic fluids, it is well known that the ``classical" description of hydrodynamics without fluctuations described above cannot reproduce 
certain low frequency behavior \cite{DESCHEPPER19741,Andreev:1978,POMEAU197563}. These effects can be incorporated into the hydrodynamic framework by adding noise to the classical equations, promoting them into stochastic equations \cite{LL-fluct,Landau:2013stat2}. Alternatively the fluctuations can be included in the hydrodynamic description as additional modes which satisfy a set of deterministic (as opposed stochastic) dynamical evolution equations that are coupled to the classical hydrodynamics  \cite{andreev1970twoliquid,Andreev:1978}. 

The effects of thermal fluctuations in relativistic hydrodynamics have also been studied, albeit more recently. The relativistic equations of motions with noise were introduced in Ref. \cite{Kapusta:2011gt}. Likewise, the evolution equations for fluctuations were first introduced in for a homogeneous, boost invariant expanding plasma (aka Bjorken flow) in Ref. \cite{Akamatsu:2017} and extended to general hydrodynamic flows in Refs. \cite{An:2019rhf,An:2019fdc}. In these works the evolution of the additional fluctuation modes are shown to be reminiscent of kinetic equations. For this reason this approach is dubbed as ``hydro-kinetics". It is worth noting that these kinetic equations describe the physics of phonons that encode fluctuations and not the microscopic degrees of freedom.  
 
In parallel, there also have been formulations of fluctuations from an effective action perspective akin to quantum field theory \cite{Kovtun:2014hpa}. Of course, the field theory in question is statistical which encapsulates not quantum but thermal fluctuations. The physics of fluctuations in a nutshell is that in addition to the microscopic diffusion processes, the conserved quantities can also be transported via hydrodynamic modes such as sounds. This backreaction of ``sound on sound" propagation alters dissipative transport both by dynamically contributing to the viscosities as well as by generating non-analytic frequency and wave-vector dependent terms in the derivative expansion \cite{Kovtun:2003,Kovtun:2011np}.  In a field theory language, these effects can be understood as loop corrections \cite{Kovtun:2012rj}. 

In this review we  summarize the recent progresses in three complimentary fronts: stochastic hydrodynamics, "hydro-kinetics", and effective field theory of hydrodynamics. The overarching goal is to present a unified picture and highlight the important progresses ranging from stochastic simulations of 3+1d hydrodynamics to more formal developments in the field. The main motivation for studying relativistic hydrodynamic fluctuations is the heavy ion collision experiments at the Relativistic Heavy Ion Collider at Brookhaven National Laboratory and Large Hadron Collider at CERN. The small droplets of 
Quark-Gluon Plasma created in these experiments can be described very well with relativistic hydrodynamics. At the same time the system size is small enough that the fluctuations can be observable. Furthermore a major goal of the heavy ion program is to search for the QCD critical point in which fluctuations play a central role. Therefore it is essential that a dynamical theory of fluctuations is developed. At the same time we will restrict the focus of this review to the theoretical developments in the study of hydrodynamic fluctuations and refer the reader to the recent reviews which focus on their role in heavy ion collisions  \cite{Bluhm:2020mpc,Du:2024wjm,Stephanov:2024mdj}.

The review is organized as follows: in Section \ref{sec:equilibrium} we briefly review equilibrium fluctuations in thermodynamics. This material is not new but included for completeness and setting the stage for what follows. Section \ref{sec:stochastic} focuses on the stochastic approach, where after going over the generalities we discuss two methods of simulating hydrodynamics: Langevin dynamics and the more recent Metropolis dynamics. The next section, Sec. \ref{sec:hydrokinetics}, is dedicated to the deterministic hydro-kinetic approach. Here we first discuss the results for the special case of boost invariant plasma, and then fluids with arbitrary flow profiles including the mathematical framework that comes with it. Then non-gaussian fluctuations which go beyond the quadratic fluctuations are analyzed. We conclude this section with the discussion of the implementation of critical dynamics in the hydro-kinetic approach. In the final section, Section \ref{sec:effective_action}, we review the effective field theory approach based on the Schwinger-Keldysh formalism and the local Kubo-Martin-Schwinger symmetry. We highlight two recent developments in this area, non classical transport coefficients and the connections to the hydro-kinetic approach via Schwinger-Dyson equations.  

%% file: equilibrium.tex
 In order to set the stage for fluctuations in a dynamically evolving background, let us first consider thermal fluctuations in a thermodynamic system at rest.  We adopt a course grained picture of the system in units of local cells, each of which is much smaller than the size of the system, hence treated as infinitesimally small. At the same time, the cells themselves are macroscopic and characterized by an equation of state. We label each cell by a coordinate $\vx$. The cells are statistically independent; therefore total entropy of the system is simply given by $\int s\,d^3x $ where $s=s(x)$ denotes the entropy of each cell, i.e. the entropy density.  The equilibrium state is determined by maximizing the entropy with the constraint that the total energy, $\int \eps \,d^3x$, and the particle number, $\int n d^3x$, are constant. As a result the equilibrium state maximizes 
  \bea
  {\cal S}=\int d^3x \left(s -\bar\beta \eps+\bar\alpha n \right)
  \label{eq:S_eff}
  \ea
  where $\bar\beta$, and $\bar\alpha$  are the Lagrange multipliers that ensures the conservation of energy and particle number. In more physical terms, they respectively correspond to the inverse temperature and chemical potential divided by temperature associated with the surrounding medium of the cell. Note that $\bar\beta$ and $\bar \alpha$ do not depend on $\vx$.  The equilibrium values of the energy and particle density are determined by maximizing ${\cal S}$: 
  \bea
  \left(\frac{\delta s}{\delta \eps}\right)_n-\bar\beta=\beta-\bar\beta=0, \qquad   \left(\frac{\delta s}{\delta n}\right)_\eps-\bar\alpha=\alpha-\bar\alpha=0
  \ea
  where $\beta=1/T,\alpha=\mu/T$.

 The fluctuations around the equilibrium state are characterized by the distribution of macrostates which is proportional to the statistical weight $e^{\cal S}$. In thermodynamics, the fluctuations around equilibrium are small and we can therefore expand this expression around the equilibrium values of the conserved densities $\eps$ and $n$. Since ${\cal S}$  is maximized in equilibrium, the linear order term in this expansion vanishes; so the fluctuations are controlled by the quadratic form composed of the second derivatives of ${\cal S}$ with respect to $\eps$ and $n$. It can be shown from elementary thermodynamic identities that the eigenmodes of this quadratic form entropy are the pressure, $p$, and specific entropy 
 \bea
 m:= \frac{s}{n}\,.
 \ea
  Therefore it is convenient switch the independent variables from $(\eps,n)$ to $(p,m)$ to diagonalize the quadratic form. In these variables, the logarithm of the statistical weight up to second order is simply 
\bea
  {\cal S}_2[p,m]= {\cal S}_0+\frac12\int d^3x\left(s_{,pp}\dpp^2+s_{,mm}\dm^2 \right)
    \ea
where the subscript 2 explicitly denotes the order we expand ${\cal S}$ in fluctuations (${\cal S}_0$ is an irrelevant constant and ${\cal S}_1$ vanishes at equilibrium), and the subscripts after a comma denote derivatives with respect to those variables, i.e. 
\bea
s_{,pp}:=\frac{\del^2s(p,m)}{\del p^2},\quad s_{,pm}=s_{,mp}:=\frac{\del^2s(p,m)}{\del p\del m},\quad s_{,mm}:=\frac{\del^2s(p,m)}{\del m^2}\,\quad s_{,ppp}:=\frac{\del^3s(p,m)}{\del p^3},\dots
\ea
The quadratic fluctuations around the equilibrium state thus can be expressed as a path integral 
\bea
\Geq_{pp}(\vx_1,\vx_2) := \av{\dpp(\vx_1)\dpp(\vx_2)}^{\rm eq}&=&\int[{\cal D} \phi ] e^{{\cal S}_2[p,m]}\,\dpp(\vx_1)\dpp(\vx_2)\nn
&=&-\left(s_{,pp}\right)^{-1} \,\delta^{(3)}(\vx_1-\vx_2)=Tw c_s^2\,\delta^{(3)}(\vx-\vx_2)\,.
\label{eq:G2eq1}
\ea
where $\phi$ is a collective notation for $(\dpp,\dm)$ and $[{\cal D}\phi]$ denotes the properly normalized path integral measure such that $\int [{\cal D}\phi]  e^{{\cal S}_2[p,m]}=1$.  Likewise for the other modes we have 
\bea
\Geq_{mm}(\vx_1,\vx_2) &:=& \av{\dm(\vx_1)\dm(\vx_2)}^{\rm eq} =-\left(s_{,mm}\right)^{-1} \,\delta^{(3)}(\vx_1-\vx_2)=\frac{c_p}{n^2} \,\delta^{(3)}(\vx_1-\vx_2) \nn
\Geq_{pm}(\vx_1,\vx_2)  &:=& \av{\dpp(\vx_1)\dm(\vx_2)}^{\rm eq}=0
\label{eq:G2eq2}
\ea
We also used the appropriate thermodynamic identities, 
\bea
s_{,pp}=-\frac{1}{Tw c_s^2},\quad s_{,mm}=-\frac{n^2}{c_p}\quad s _{,mp}=0
\label{eq:spp_and_smm}
\ea      
with $w=\eps+p$ being the enthalpy density,  $c_s^2=(\del p/\del \eps)_m$, the speed of sound and  $c_p=Tn(\del m/\del T)_p$, the heat capacity at constant pressure.
Note that the negativity of the terms in Eq. \eqref{eq:spp_and_smm} are dictated by the second law, namely the small deviations from the equilibrium decrease the entropy. 
It is worth pointing out that the equilibrium correlators given in Eqs. \eqref{eq:G2eq1} and \eqref{eq:G2eq2} are local and their magnitude is completely determined by thermodynamics via the equation of state. Of course, what is meant here by ``local" is in the context of a course grained macroscopic system where each infinitesimally small cell is still a macroscopic 

The higher-order fluctuations are encoded in multi-point functions such as 
\bea
\Geq(\vx_1,\vx_2,\vx_3)& :=& \av{\phi(\vx_1)\phi(\vx_2)\phi(\vx_3)}^{\rm eq},
\\
\Geq(\vx_1,\vx_2,\vx_3,\vx_4)& :=& \av{\phi(\vx_1)\phi(\vx_2)\phi(\vx_3)\phi(\vx_4)}^{\rm eq}\,.
\ea 
They can be calculated in a straightforward way by expanding the entropy into the appropriate order and evaluating the path integral. For instance, in order to calculate the three point function, $\av{\delta m(\vx) \delta m(\vy)\delta m(\vz) }$, we need to expand the entropy functional to cubic order:
 \bea
   {\cal S}_3= {\cal S}_2+  \int d^3x \left(\frac16 s_{,mmm} \delta m^3 +\dots\right)
 \ea
where $\dots$ refer to all the other derivatives of $s$ that include $p$ which do not appear in this correlation function. Evaluating the path integral,
 \bea
  \av{\delta m(\vx_1)\delta m(\vx_2)\delta m(\vx_3)}=\int [{\cal D}   \phi] e^{{\cal S}_2} \int d^3x^\prime \delta m(\vx_1)\delta m(\vx_2)\delta m(\vx_3)   \left(\frac16s_{,mmm}\right)\delta m({\bf x^\prime})\delta m({\bf x^\prime})\delta m({\bf x^\prime})\,,
 \ea
and  performing the Wick contractions, each of which brings a factor of  the ``propagator", $-(s_{mm})^{-1}$, leads to 
 \bea
   \av{\delta m(\vx_1)\delta m(\vx_2)\delta m(\vx_3)}&=&-\frac{ s_{,mmm}}{(s_{,mm})^3 } \delta^{(3)}(\vx_1-\vx_2)\,\delta^{(3)}(\vx_1-\vx_3)\\
   &=&\left(\frac{c_p}{n^2}\right)^2\left[\left(\ln\frac{c_pw^2}{T^2n^6}\right)_{,m}-\frac{2 T n}{ w}+\frac{n}{c_p}\right]\,\delta^{(3)}(\vx_1-\vx_2)\,\delta^{(3)}(\vx_1-\vx_3)
 \ea
 Here we used various thermodynamic identities to simplify the expression. More details can be found in Ref. \cite{An:2022jgc}.  
 
\begin{figure}[h]
\center
\includegraphics[scale=0.5]{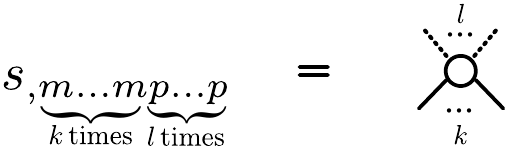}
\caption{In the diagrammatic language, three and higher derivatives of entropy are represented as vertices. Solid and dashed lines represent $p$ and $m$ respectively}
 \label{fig:vertex}
\end{figure}

\begin{figure}[h]
\center
\includegraphics[scale=0.5]{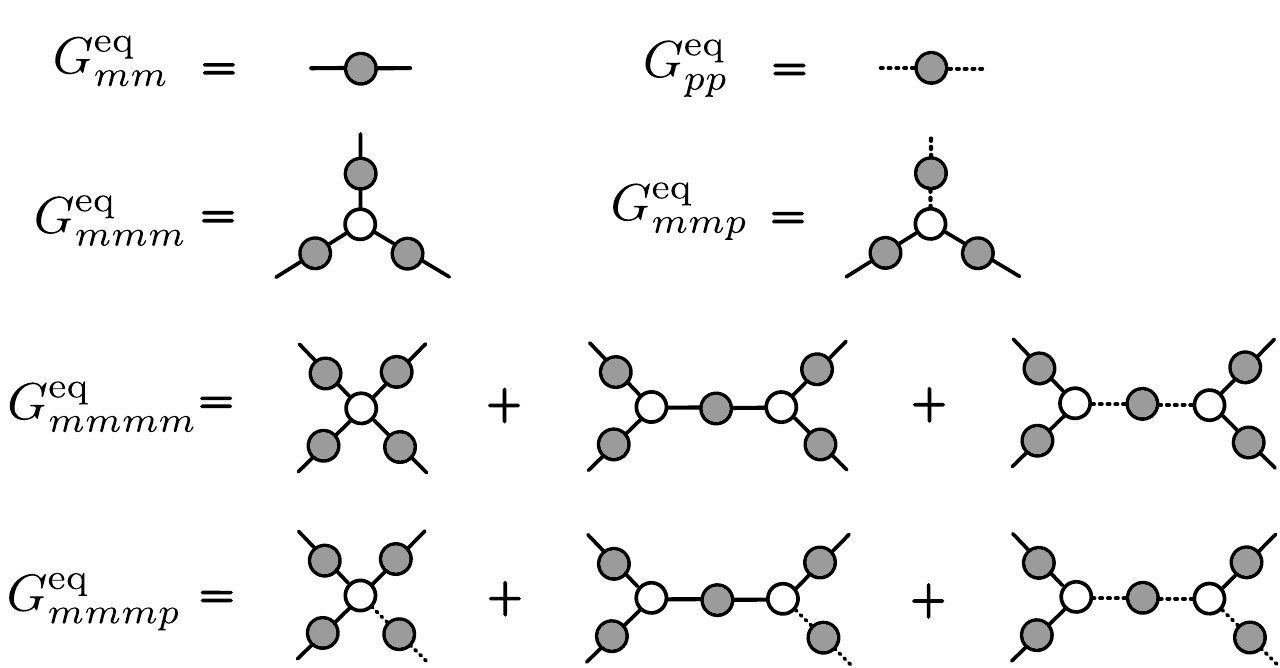}
\caption{The diagrammatic representation of equilibrium correlation functions.}
 \label{fig:diagrams_eq}
 \end{figure}

 This pattern can be generalized to an arbitrary $n-$point function in a straightforward way. A convenient way to represent the higher-point correlators is through diagrams \`a la quantum field theory. In the diagrammatic language, three and higher order derivatives of the entropy are represented as vertices shown in Fig.~\ref{fig:vertex}
The two point functions $G^{\rm eq}_{mm}$ and $G^{\rm eq}_{pp}$ are represented as solid circles with two solid and dashed legs respectively. Any higher point function is  expressed as the sum of all possible, connected tree level diagrams that can be drawn by joining the vertices with the two point functions. Some examples are shown in Fig.~\ref{fig:diagrams_eq}.
  
  Let us now consider a relativistic fluid. The conserved energy-momentum tensor of a relativistic hydrodynamic system can be expressed as\footnote{In this review we work in the Landau frame and our metric signature is $-+++$. For a recent review of hydrodynamic frames we refer the reader to \cite{Bhattacharyya:2024ohn}. }
   \bea
 T^{\mu\nu}=\eps u^\mu u^\nu+p\bD^{\mu\nu}=wu^\mu u^\nu+p g^{\mu\nu}
 \ea 
where $u^\mu$ is the normalized four velocity of the fluid with  $u^2=-1$ and $\Delta^{\mu\nu}=g^{\mu\nu}+u^\mu u^\nu$ is the transverse projection operator perpendicular to $u$. The energy-momentum tensor exhibits thermal fluctuations. Just like the static case discussed above, the equilibrium fluctuations are governed by the conserved quantities. In this case in addition to the energy density, momentum density fluctuates as well. In the lab frame, they take the form
\bea
T^{00}&=&\gamma^2 w-p\\
T^{0i}&=&\gamma w v^i 
\ea
where the four-velocity is $u=\gamma(1,{\bf v})$ with $\gamma=(1-{\bf v}^2)^{-1/2}$ being the usual Lorentz factor. 
Likewise the charge density also picks up a Lorentz factor, i.e. $J^0= \gamma n$. The logarithm of the statistical weight for the fluctuations of the fluid is then 
\bea
{\cal S}=\int d^3x \left(\gamma s -\bar\beta (\gamma^2 w-p)+ \gamma \bar\alpha n  \right)
\ea
For simplicity, let us consider the fluctuations around a static background such that the average velocity vanishes; ${\bf \bar v}=0$. Repeating the same steps as above, we expand ${\cal S}$ to quadratic order in velocity. From the expansion of the Lorentz factor,
\bea
\gamma\approx 1-\frac12\delta v^2,
\ea
we obtain
 \bea
  {\cal S}_2[p,m,{\bf v}]={\cal S}_0+ \frac 12 \int d^3x \left(s_{,pp}\dpp^2+s_{,mm}\dm^2-\beta w \delta v^2 \right)\,.
 \ea
The fluctuations of velocity therefore follows from the gaussian path integral as above:
 \bea
 \av{\delta v_i(\vx_1)\delta v_j(\vx_2)}&=&\frac{T}{w}\delta_{ij}  \,\delta^{(3)}(\vx_1-\vx_2)
 \ea
The higher order correlators of velocity can also be obtained in a diagrammatic way with the addition of a vector values leg that represents $\bf v$.

%% file: stochastic.tex
\subsection{General formulation}
\label{sec:stc_general}
In a majority of cases of interest, such as heavy ion collisions, the  fluctuations occur in an evolving medium, necessitating a theory of \textit{dynamics} of fluctuations. The underlying physics can be understood as follows. Hydrodynamics describes a macroscopic system whose parts are not in complete, global thermal equilibrium. The  variations of the thermodynamic variables over a spatial scale $L$ is characterized in terms of a gradient expansion. The relevant degrees of freedom in this expansion are densities of the conserved quantities, such as energy-momentum and particle number. This is because, in the timescale of the hydrodynamic evolution, $\tau \sim L/c_s$, the rest of the degrees of freedom are assumed to be already thermalized; as we eluded to in the Introduction. The thermalization of the hydrodynamic modes occurs in a longer time scale as it happens through diffusion. Therefore in the timescale $\tau$, thermodynamic equilibrium can only be established in a spatial scale that can be reached via diffusion: 
\bea
\ell_{\rm f}\sim \sqrt{\gamma_d \tau}\sim\sqrt{\gamma_d L/c_s}
\ea
where $\gamma_d$ is the appropriate diffusion coefficient, proportional to a combination of viscosities and charge diffusion constant depending on the mode in question.  For example, for shear diffusion it is given by $\gamma_\eta=\eta/(\epsilon+p)$ where $\eta$ is the shear viscosity. Here the subscript f stands for ``fluctuation", reminding that this is the characteristic scale where fluctuations start to fall out of equilibrium. 

Through this physics of equilibration, a new hierarchy of scales emerges. The short wavelength fluctuations with $\lambda \ll \ell_{\rm f}$ are thermalized and therefore are equal to the equilibrium values determined entirely by the equation of state as discussed in Sec.~\ref{sec:equilibrium}. These contribution of these short-distance modes to the fluid motion is local and can be absorbed into the equation of state. We will elaborate on this procedure in more detail below.  
On the other hand, longer wavelength fluctuations with $ L\gg \lambda \gtrsim \ell_{\rm f} $ are not in thermal equilibrium and exhibit non-trivial dynamics. At the same time, they occur in shorter scales compared to the background flow. 
These non-equilibrium modes contribute to the fluid motion both locally and non-locally. The local part constitutes a dynamical contribution to dissipative transport and can be absorbed into the viscous terms. Whereas the non-local contribution cannot be described in terms of classical hydrodynamics and lead to non-analytic terms in the derivative expansion which encode the famous long-time tails \cite{DESCHEPPER19741,Arnold:1997gh,Kovtun:2003,Kovtun:2011np}.

\begin{figure}[h]
\center
\includegraphics[scale=0.5]{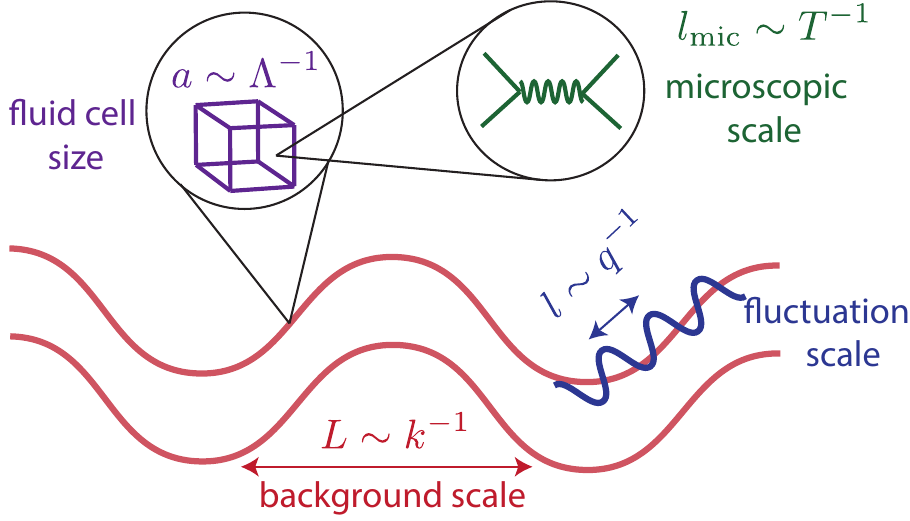}
\caption{A depiction of different scales in a hydrodynamic system: $k \ll q < a < \lmic$. Figure credit: Xin An.}
\label{fig:scales}
\end{figure}

 In Fourier space, the characteristic wave-vector, $q\sim1/\ell_{\rm f}$, associated with non-equilibrium fluctuations lies in the range 
 \bea
 k \ll q\sim \sqrt{k/\gamma_d} \ll T
 \ea
  where $k\sim1/k$ is the typical wave-vector of the background hydrodynamic motion. For future reference we also denote the typical hydrodynamic cell size as $a\sim1/\Lambda$ and the typical microscopic scale as $\lmic\sim1/T$. As mentioned,  each hydrodynamic cell is assumed to be a macroscopic system (in the sense that $a\gg\lmic$) in local thermal equilibrium so that the equation of state is defined locally. This hierarchy of scales is illustration in Fig.~\ref{fig:scales}.

A standard approach to incorporating the dynamics of fluctuations in the framework of hydrodynamics goes back to the work of Landau-Lifshitz where the hydrodynamic equations are promoted to stochastic differential equations \cite{Landau:2013stat2}. The fluctuations manifest themselves as random, instantaneous kicks that contribute to the fluxes.  Their magnitude is proportional to the viscosities per the fluctuation-dissipation theorem. This stochastic approach was generalized to relativistic hydrodynamics in \cite{Kapusta:2011gt}.
The starting point is the conservation of energy-momentum and charge current (for cases with conserved charge) as usual:
\bea
\partial_\mu\sT^{\mu\nu}=0  \\
\partial_\mu\sJ^{\mu}=0 \,.
\label{eq:sth_cons}
\ea
But now the components of the energy-momentum tensor and charge current are \textit{stochastic} variables that exhibit fluctuations. Following the notation introduced in \cite{An:2019rhf} we denote stochastic variables with the ``breve"  accent, i.e. $\,\,\sth\,\,$. Just like the equilibrium case, the fluctuations stem from the fact that each fluid cell itself is a macroscopic system with finite volume and therefore exhibit usual thermal fluctuations. Let us separate the conserved currents into two parts: the constitutive part and random noise,
\bea
\sT^{\mu\nu}&=&\sT_{c}^{\mu\nu}+\sS^{\mu\nu}\\
\sJ^{\mu}&=&\sJ_{c}^{\mu}+\sI^\mu
\ea
Here the subscript ``c" refer to the constitutive currents where the energy-momentum tensor and the charge current are expressed as functionals of some set of fluid variables like energy density and fluid velocity which, at this point, is not specified. The second terms $\sS^{\mu\nu}$ and $\sI^\mu$ denote the random noise that encodes the thermal fluctuations. In the course grained picture, where the fluid cells are considered to be infinitesimally small, the noise is local, Gaussian and point-wise correlated with a magnitude proportional to the dissipative coefficients \cite{Kapusta:2011gt}:
\bea
        \av{\sS^{\mu\nu}(x)\sS^{\lambda\kappa}(x^\prime)}&=& 2T\left[ \eta\, (\bD^{\mu\kappa}\bD^{\nu\lambda}+\bD^{\mu\lambda}\bD^{\nu\kappa})+\left(\zeta-\frac{2}{3} \eta\right) \bD^{\mu\nu}\bD^{\lambda\kappa} \right] \delta^{(4)}(x-x^\prime)\,, 
         \label{eq:gaussian_noise_Pi}
        \\
   \av{\sI^{\mu}(x)\sI^{\nu}(x^\prime)}&=& 2\lambda \bD^{\mu\nu}\delta^{(4)}(x-x^\prime)\,.
 \label{eq:gaussian_noise_J}
\ea 
Here $\eta$, $\zeta$ are the shear and bulk viscosities and $\lambda= T \sigma$ with $\sigma$ being the charge conductivity. We will also make use of the diffusive coefficients
\bea
\gL=\frac{1}{w}\left(\zeta+\frac{4}{3}\eta\right)+ \lambda c_s^2 Tw\left(\left(\pd{\alpha}{p}\right)_m\right)^2,\quad  \geta=\frac{\eta}{w} , \quad\gamma_\lambda=\frac{\kappa}{c_p}
\ea
where
\bea
\kappa=\lambda\left(\frac{w}{T n}\right)^2
\ea
is the thermal conductivity. 
The projection operator 
\bea
\bD^{\mu\nu}=g^{\mu\nu}+u^\mu u^\nu
\ea
ensure that the noise is spatial in the local rest frame of the fluid. As explained earlier, the appearance of the transport coefficients is a result of the fluctuation dissipation theorem. The non-Gaussian terms are suppressed by the volume as a consequence of the central limit theorem and therefore are justifiably neglected in the thermodynamic limit. To be more explicit, the number of uncorrelated cells in a volume $L^3$ is proportional to $(L/\xi)^3$ which in the hydrodynamic limit is a large number. According to the central limit theorem, the magnitude of fluctuations scale as $1/\sqrt{N}\sim (\xi/L)^{3/2}$ and the deviations from gaussianity are further suppressed by higher powers of $1/N$. Moreover, away from a critical point where the correlation length becomes large, the correlation length is determined by the microscopic length scale $\xi\sim\lmic\sim1/T$.   

We can choose the independent fluid variables variables to be the pressure  $\sth p$, specific entropy density $\sth m$ and the four-velocity of the fluid $\sth u^\mu$ which is normalized, i.e. $\sth u^2=-1$. In the Landau frame $\sth u^\mu$ is defined as the time-like eigenvector of the energy-momentum tensor:
\bea
 \sT^{\mu\nu}\sth u_\nu =\sth \eps \sth u^\mu
\ea
and similarly the charge current satisfies 
\bea
\sth u_\mu \sJ^\mu=\sth n 
\ea
In what follows, we will denote functions of the primary stochastic variables, $\sth p$, $\sth m$ and $\sth u^\mu$, with an accent as well. For example, the fluctuating energy density will be denoted as $\sth \eps:=\eps(\sth p, \sth m)$ whose functional form is determined by the equation of state. We also use the umbrella notation $\sth \phi=(\sth m,\sth p,\sth u^\mu)$ to compactly denote the set of primary hydrodynamic variables. The constitutive relations in the Landau frame are
\bea
\sth T_{c}^{\mu\nu} &:=& T_{c}^{\mu\nu}[\sth\phi] =\sth\eps\sth u^\mu\sth u^\nu+\sth p\sth\Delta^{\mu\nu} +\sth \Pi^{\mu\nu}
\label{eq:constitutive_Pi}
\\
\sJ_{c}^\mu&:=&J_{c}^\mu [\sth\phi] = \sth n \sth u^\mu+\sJ_D^\mu
\label{eq:constitutive_J}
\ea
where $\Pi$ and $J_D$ denote the dissipative terms which are higher order gradients of fluid variables. To leading order  in derivatives they are
\bea
\Pi^{\mu\nu}&=&-\eta \sigma^{\mu\nu}-\zeta\Delta^{\mu\nu}\theta
\label{eq:dissipative_Pi}
\\
J_D^\mu&=&-\lambda \bD^{\mu\nu}\del_\nu \alpha\,.
\label{eq:dissipative_J}
\ea
The velocity gradients are organized respectively into traceless shear and bulk contributions as usual
\bea
  \sigma^{\mu\nu}\equiv \Delta^{\mu\alpha}\Delta^{\nu\beta}(\partial_\alpha u_\beta+\partial_\beta u_\alpha)-\frac{2}{3} \Delta^{\mu\nu}\del\cdot u\,, \quad
        \theta\equiv\theta^\mu_\mu=\partial\cdot u \,.
\ea
To sum up, with the constitutive relations given in Eqs. \eqref{eq:constitutive_Pi}, \eqref{eq:constitutive_J} and the noise correlators given in Eqs.  \eqref{eq:gaussian_noise_Pi} and \eqref{eq:gaussian_noise_J}, the conservation equations form a closed set of stochastic partial differential equations which define relativistic hydrodynamics with fluctuations. The rest of this section is dedicated to the overview of two main approaches to directly solving them numerically. In the next section, Sec. \ref{sec:hydrokinetics}, we will discuss an alternative approach which involves converting these stochastic equations into a system of deterministic equations for the correlation functions of the conserved quantities.  

\subsection{Langevin dynamics}

With the exception of a few special examples (e.g. \cite{Bjorken:1982qr,Gubser:2010ze}), it is generally not possible to express the solutions to the equations for relativistic fluid dynamics analytically due to the complicated non-linear nature of the equations. A common numerical scheme to solve them, in a nutshell, is to discretize space-time and evaluate the gradients as finite derivatives. Over the years, there has been remarkable advances in constructing various grid algorithms that minimize numerical instabilities. The state-of-the-art algorithms can handle 3+1 dimensional fluid dynamics with small viscosities and relatively large gradients. For publicly available codes of various numerical implementations of 3+1 d relativistic hydrodynamics see e.g. MUSIC \cite{Schenke:2010nt} , iEBE VISHNU \cite{Shen:2014vra}, vHLLE \cite{Karpenko:2013wva}, ECHO-QGP\cite{DelZanna:2013eua}, superSONIC \cite{supersonic}.    

A straightforward way to incorporate fluctuations in these numerical methods is to add random noise to the fluxes at each time step in the fluid evolution. The resulting set of equations is a generalization of the famous Langevin equation which describes random walk of a particle in the presence of a drift force. In fluid dynamics, each grid site represents a degree of freedom which acquires random kicks throughout the Langevin evolution and interacts with the neighboring sites nonlinearly. In this analogy, the drift force represents the hydrodynamic forces and the random kicks represent the thermal noise. Loosely speaking, going from single particle Brownian motion described by the Langevin equation to stochastic hydrodynamics is akin to going from single particle quantum mechanics to quantum field theory.

In practical implementations of thermal noise, one should consider various instabilities that arise upon numerical implementation of relativistic Navier-Stokes equations up to first order gradients \cite{Hiscock:1985zz,Geroch:1990bw,Geroch:1995bx}.  A common way of dealing with these pathologies regarding stability as well as causality is to promote the viscous contribution to the stress-energy tensor  $\Pi^{\mu\nu}$ to a dynamical field which relaxes to the equilibrium value with an empirically chosen relaxation time, $\tau_\Pi$, which is of the order as the microscopic relaxation time $\lmic/c_s$:
\bea
u\cdot \del \Pi^{\mu\nu}=-{1\over\tau_\Pi}\left( \Pi^{\mu\nu}+\eta \sigma^{\mu\nu}+\zeta \theta \Delta^{\mu\nu}\right)
\ea
This goes under the name Israel-Stewart theory \cite{is}. When fluctuations are taken into account, it is convenient to evolve the noise through a similar relaxation dynamics:
\bea
u\cdot \del \sth S^{\mu\nu}=-{1\over\tau_\Pi}\left( \sth S^{\mu\nu}-\sth S^{\mu\nu}_{0}\right)
\ea
where it is now $S_0^{\mu\nu}$ that satisfies the point-wise Gaussian correlation given in Eq.~\eqref{eq:gaussian_noise_Pi} \cite{Young:2013fka,young_kapusta_gale,Bluhm:2018plm}. In practice, there are various other implementations of second order-order hydrodynamics \cite{Baier:2007ix} that that can be used to cure the numerical instabilities of first-order hydrodynamics \cite{Bemfica:2017wps,Kovtun:2019hdm} some of which have been extended to include fluctuations \cite{Kumar:2013twa,Murase:2013tma,Fujii:2023iwt}. With all these advances, the state-of-the-art stochastic  Langevin-hydrodynamics have been used to simulate 3+1 dimensional hydrodynamics \cite{Singh:2018dpk,young_kapusta_gale,SAKAI2017445,MURASE2016276} as well as critical diffusion models \cite{Sakaida:2017rtj,Nahrgang:2018afz,Bluhm:2019yfb,Kitazawa:2020kvc}. For an extensive review of these advances that focuses on the applications to heavy ion collisions we refer the reader to Ref. \cite{Bluhm:2020mpc}. 

In addition to the issues regarding causality and stability which are already present in non-fluctuating hydrodynamics and can be addressed by introducing relaxation dynamics, there are various other challenges in the Langevin approach that are particularly due to the implementation of the noise. The first technical issue stems from the locality of the noise. Since noise at different space-time points is uncorrelated, the configurations generated by the Langevin evolution have very large gradients that are challenging for any type of numerical partial differential equation solver. 

Furthermore, the magnitude of the noise at every point is large compared to the background (drift) terms due to the delta function in Eqs. \eqref{eq:gaussian_noise_Pi} and \eqref{eq:gaussian_noise_J}. In practice the singularity of the delta function in the coincident limit is regularized and is inversely proportional to the grid spacing, $\delta^{(4)} (x-x^\prime)\sim1/(\Delta t\times \Dx^3)$, which is smaller than any scale in the problem. Again, from the noise correlator, Eqs. \eqref{eq:gaussian_noise_Pi} and \eqref{eq:gaussian_noise_J}. we can that the magnitude of the noise terms in the evolution scale as 
\bea
\sS^{\mu\nu}, \sI^\mu \sim\frac{1}{\sqrt{\Delta t \Dx^3}}
\ea
whereas the drift terms are ${\cal O}(1)$. Therefore upon averaging over noise, the physical observables acquire a dependence  on the lattice spacing. In our analogy with quantum field theory, these are the typical short-distance (UV) singularities. 
 In other words, they explicitly depend on the UV cutoff scale which is the inverse lattice spacing. In principle, the UV sensitive terms renormalize the equation of state and the transport coefficients. We can express the contribution of the fluctuations to the average energy momentum tensor and current as 
\bea
\av{\sT^{\mu\nu}}&=&T_c^{\mu\nu}[\av{\sth\phi}]+\Delta T^{\mu\nu} \\
\av{\sJ^{\mu}}&=&J_c^{\mu}[\av{\sth\phi}]+\Delta J^{\mu} 
\ea  
Here the first terms in each equation correspond to the contribution of the average fluid flow to the conserved fluxes which, in general, differ from the true noise average quantities due to the nonlinear dependence of $T^{\mu\nu}$ and $J^\mu$ on $\phi$. This difference is due to fluctuations. Schematically the UV dependent part is of the form
\bea
\Delta T^{\mu\nu} \sim a T \Lambda^3+  b \frac{T \Lambda}{\gamma_d} \del u+ \dots   \\
\Delta J^{\mu} \sim c T \Lambda^3+  d \frac{T \Lambda}{\gamma_d} \del \alpha+ \dots  \,.
\label{eq:renormalization}
\ea
with the appropriate tensor structure whose exact form can be found in Ref. \cite{An:2019fdc}. Here $\gamma_d$ generically refers to the appropriate diffusion coefficient proportional to the viscosities and the conductivity. The dots refer to the finite terms which are in general non-analytic in the frequency and wave-vector and correspond to the long-time tails. The first terms that depend on $\Lambda^3$ are due to equilibrium fluctuations discussed in Sec.~\ref{sec:equilibrium}, and they can be canceled by appropriate counter-terms for $\eps$ and $n$. In other words, they renormalize the \textit{bare} values of $\eps$ and $n$.  The second terms are due to the local contribution of non-equilibrium fluctuations to the dissipative terms and they renormalize the viscosities and conductivity. Note that remarkably they are inversely proportional with the bare values of the transport coefficients as pointed out in Ref. \cite{Kovtun:2011np}. We will elaborate on renormalization and provide more precise form of these schematic formulas in Sec.~\ref{sec:hydrokinetics}. 

Even though this renormalization procedure is conceptually straightforward, it is practically challenging to implement in the Langevin approach. The bare viscosities enter as inputs in the numerical simulations. However due to the dynamical contribution of the fluctuations the physical viscosities differ from the bare values. Therefore the bare values have to be fine-tuned based on the lattice spacing such that the dynamical evolution reproduces the correct physical values. This can be a numerically challenging problem. In addition to the difficulties with renormalization, the large noise can also lead to unphysical configurations where certain sites have negative energy or particle number densities. In order to mitigate these issues certain methods have been developed, such as including smearing the noise with some coarse graining procedure \cite{MURASE2016276,Bluhm:2018plm}, or applying a hard cut-off filter which removes fluctuation modes whose wave-vectors are greater than some cutoff scale $\Lambda_{cutoff}$ with $k < \Lambda_{cutoff}  < \Lambda$ where $k$ denotes the typical wave-vector of the average fluid flow \cite{SINGH2019319}.

Another potential difficulty of the Langevin approach is the so-called ``multiplicative noise" issue. This issue stems from the fact that the magnitude of the noise in the Langevin-hydro equations depends on the stochastic variable $\sth\phi$. This can be seen directly from Eqs. \eqref{eq:gaussian_noise_Pi} and \eqref{eq:gaussian_noise_J} where the noise correlator depends on the transport coefficients which are functions of the stochastic variables, energy and number densities, as well as the fluid velocity, another stochastic variable. This dependence is the source of a famous ambiguity that arises when the system is discretized in time. Let us schematically write the Langevin equation with discrete time evolution as 
\bea
\sth\phi_{t+\Delta t}=\sth\phi_{t}+ \Delta t F[\sth\phi_t]  + \sqrt{2\Delta t} H[\sth\phi^\prime_t]  \xi_t
\ea   
where for simplicity we suppressed all the index structure as well as the spatial grid dependence in the notation. The $F$ term denotes the drift/drag terms. In hydrodynamics it represents the contribution from the constitutive currents (both ideal and dissipative). The $\xi_t$ in second term is the noise sampled from a gaussian with zero mean and unit variance. The function $H$ determines the magnitude of the noise and is fixed by Eqs. \eqref{eq:gaussian_noise_Pi} and \eqref{eq:gaussian_noise_J}. 

The key point is that different discretization prescriptions where $H$ is evaluated at different time points, for instance at $\sth\phi_t$ versus at $\sth\phi_{t+\Delta t}$, lead to different answers. Two common choices are $\sth\phi^\prime_t:=\sth\phi_t$ and  $\sth\phi^\prime_t:=(\sth\phi_t+\sth\phi_{t+\Delta t})/2$, respectively known as It$\bar {\rm o}$ and Stratonovich prescriptions. In general, there is a continuum of different choices defined by a linear combination of $\sth\phi_t$ and $\sth\phi_{t+\Delta t}$.
This choice matters only for $H$ and not for $F$ because for $F$, the difference between prescriptions is an ${\cal O}(\Delta t)$ effect and therefore vanishes in the continuum limit. However due to the fact that the noise term is proportional to  $ \sqrt{2\Delta t} $, the difference between prescriptions in the noise term survives the continuum limit. In fact, the difference between discretization prescriptions can be recast as a shift in the drift term $F$ \cite{Arnold:1999va,Arnold:1999uza}. 

Therefore, strictly speaking, without a prescription for discretization, the Langevin-hydro equations written generally as in Eq.~\eqref{eq:sth_cons} are not well-defined. In the context of fluctuations in relativistic hydrodynamics, this ambiguity has been discussed in detail in \cite{Young:2013fka}. It is now understood that there is no ``correct choice" for the prescription, but for any choice one makes, the drift term has to be adjusted accordingly such that the system obeys the correct equilibrium distribution, as astutely pointed out in \cite{Arnold:1999va,Arnold:1999uza}.  At the same time, doing so practically can be numerically challenging for a similar reason regarding  renormalization. In a nutshell, it requires the drift term to be fine tuned such that it correctly reproduces the equilibrium distribution which can only be checked after running the simulation.

\subsection{Metropolis dynamics}

The Langevin approach to incorporating fluctuations in hydrodynamics is conceptually straightforward; at each step of the evolution the fluid is subject to random kicks whose distribution is determined by the fluctuation-dissipation theorem. At the same time, the practical implementations of this approach are challenging due to the infinite noise and multiplicative noise problems as explained in the previous section. To mitigate these problems, an alternative method for incorporating fluctuations in fluid dynamics based on Metropolis algorithm has been introduced \cite{Florio:2021jlx,Schaefer:2022bfm}. 

The Metropolis algorithm is a ubiquitous Markov chain Monte-Carlo method, typically used to study statistical systems  where a direct calculation of physical observables (i.e. correlation functions of fields) is not possible. Notable famous examples include strongly interacting quantum (or statistical) field theories such as QCD and 3d Ising model \cite{ZinnJustin:2002ru}. In the Monte-Carlo framework, instead of a direct calculation of physical observables, one samples the field space by generating a random sequence of field configurations that is distributed with respect to the probability density $e^{-I[\phi]}$. Here $I$ is the action functional associated with the system in question\footnote{We reserve the letter $S$ to denote the entropy.}, and $\phi$ generically denotes the underlying field. The physical correlation functions then are approximated by the correlation functions computed over the sample. Assuming that the system is ergodic and thermalized, this approximation converges to the physical value as $1/\sqrt{N}$ where $N$ is the number of field configurations in the sample.   

The algorithm in a nutshell has two steps: 1) based on the existing field configuration, $\phi$, propose a new one, say $\phi^{new}$, and 2) either accept or reject this configuration with the acceptance probability min$(1,e^{-I[\phi_{new}]+I[\phi]})$. This procedure guarantees that the field configurations in the Markov chain are distributed with respect to  $e^{-I[\phi]}$.

The application of this method to stochastic systems stems from the fact that the equilibrium fluctuations originate from the distribution $e^{S[\phi]}$ where $S$ is the entropy, as explained in Sec.~\ref{sec:equilibrium}. Furthermore, according to the fluctuation dissipation theorem, the dissipative forces are related to the entropy as well. For example, the  in the nonlinear diffusion problem the constitutive current is purely dissipative and can be written as
\bea
{\bf J}_c=-\lambda {\bf \nabla} \alpha= \lambda {\bf \nabla} \frac{\delta S[n]}{\delta n(x)}\,.
\ea
This equation follows from the thermodynamic relation $\del S/\del N=-\mu/T=-\alpha$. As a result, the Markov chain created by the Metropolis algorithm captures the dynamics of the thermalization process towards equilibrium, just like the Langevin dynamics. However, its implementation is different from Langevin method, which makes Metropolis approach practically appealing as we detail below.  

Before detailing the application of the Metropolis algorithm in stochastic fluid dynamics, we pause to compare and contrast with its application in lattice field theory, for example lattice QCD \cite{Montvay:1994cy}. In field theory, Metropolis algorithm is used to compute the equilibrium properties of the system, such as the equation of state or static properties such as Euclidean time correlation functions, from which quantities like hadron spectra can be obtained. Here, the ``Metropolis time" parameterizes the evolution of field configurations in the Markov chain and is not physical. In order to obtain a sample that represents the equilibrium state, one needs to make sure the system is thermalized. This means the total Metropolis time has to be significantly larger than the thermalization time. Furthermore, in order to optimize the computational resources, it is also to beneficial to engineer algorithms such that the system equilibrates as quickly as possible and the successive configurations in the Markov chain have little autocorrelation. This can be achieved via modifying the proposal and accept/reject steps. There are many different methods, such has Hybrid Monte-Carlo algorithm \cite{Brooks_2011}, at one's disposal for these purposes.  In stochastic fluids, similarly the Metropolis algorithm ensures that the fluid locally equilibrates to a state where the hydrodynamic degrees of freedom are sampled from the equilibrium distribution, $e^{S}$. However, in contrast to field theory, the Metropolis time in this case \textit{is} the physical time and the evolution of the Markov chain corresponds to the physical thermalization process which we are interested in studying. One might summarize this difference as follows. In lattice field theory, we are interested in getting to our destination of equilibrium as quickly as possible, whereas in stochastic fluids, it is the trip that carries the interesting physics.  

 The Metropolis algorithm for stochastic fluids has three main steps: \\
 1) Proposal: Given a configuration $\phi_{\vx,t}$, a new configuration $\phi^{new}_{\vx,t}=\phi_{\vx,t}+{\rm noise}_{\vx, t}$ is proposed. The noise satisfies the fluctuation dissipation and conservation relations.  
 \\
 2) Accept/reject: The proposed field is accepted (meaning $\phi_{\vx,t}$ is updated to $\phi^{new}_{\vx,t}$) with the probability min$(1,e^{S[\phi^{new}]-S[\phi]})$. If the proposal is rejected, $\phi_{\vx,t}$ remains unchanged. Steps 1 and 2 are repeated over the entire lattice.  
 \\
 3) Deterministic update:  The updated field $\phi_{\vx,t}$ is evolved into $\phi_{\vx,t+\Delta t}$ by the deterministic and \textit{non-dissipative} (advective / ideal hydro) equations. At this step, different types of integrators developed for ideal hydrodynamics can be used to maintain conservation and stability as much as possible in the discretized form \cite{Kurganov:2000ovy,Zanna:2002qr}.

Notably, in this algorithm,  \textit{there is no dissipative step}. Dissipation is generated dynamically via the proposal-accept/reject steps, in other words, via fluctuations. Let us illustrate how this works in the nonlinear diffusion case where there is no advection, i.e. there is no third step. The only degree of freedom in the problem is charge density, $n$. The constitutive current is given by
\bea
J_c^\mu=(n,-\lambda {\bf \nabla \alpha})
\ea
where $\lambda$ and $\alpha$ are, in general, functions of $n$. We can consider a simple discretization where $\vx$ is defined on a cubic lattice with lattice spacing $a$, and the charge $n_\vx$ sits on the sites of this lattice. The time evolution is also discretized with a spacing $\Delta t$.
\begin{figure}[h]
\center
\includegraphics[scale=0.7]{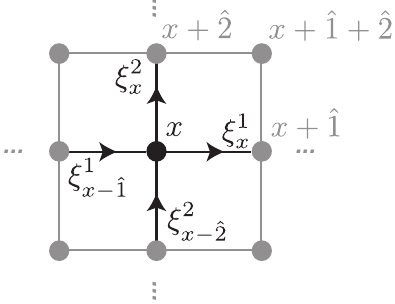}
\caption{A two dimensional depiction of the Metropolis update for diffusion. The charge density, $n_\vx$ is updated locally by a random flux. The arrows denote the relative sign of the noise contribution to charge given in Eq.~\eqref{eq:n_metropolis_update} and the equality of ingoing and outgoing arrows at every site is the manifestation of charge conservation.}
\label{fig:lattice}
\end{figure}
The random charge flow is generated by the proposal 
\bea
J^{i,new}_{\vx}=\sqrt{\frac{2\lambda}{\Delta t \Delta V}} \,\xi_{\vx}^i
\ea
where the random variable $\xi_{\vx}^i$ is sampled from a normalized gaussian distribution:
\bea
\av{\xi_{\vx}^i \,\xi_{\vy}^j}=\delta_{\vx\vy}\delta^{ij}
\label{eq:met_noise}
\ea
The factor $\Delta V$ denotes the effective ``noise volume", i.e. the volume of each cell which fluctuates independently. It is a microscopic scale which could be taken as $a^3$ or can be smeared out to a larger (yet still microscopic) volume to tame the short-distance divergences, similar to what is practiced in Langevin simulations. As long as the renormalization is carried out properly, the results must be independent of the choice of $\Delta V$. The charge is updated in accordance with the conservation law. Namely the proposed charge is
\bea
n^{new}_{\vx}=n_\vx-\frac{\sqrt{2\lambda\Delta t/ \Delta V}}{a} \sum_{i\in\{1,2,3\}} \left(\xi_\vx^i-\xi_{\vx-\hat i}^i\right)\,.
\label{eq:n_metropolis_update}
\ea
This form ensures that the change in the charge at every lattice site is balanced by the incoming and outgoing fluxes of current as depicted in Fig.~\ref{fig:lattice}). It is useful to visualize the random current to live on the links. To maintain conservation, all the links that are in contact with $\vx$ contribute to the local update.
Notice that there is no deterministic dissipative contribution to the update. Following the logic in Ref. \cite{Basar:2024qxd} we now show how dissipation is generated via noise by calculating the average current.
There are two factors that determine the overall probability distribution of the configurations: the probability of a configuration to be proposed at the first place, $P_{\rm prop.}$ and the probability of it being accepted, $P_{\rm accept}$. The average current is given by the weighted average over all possible configurations:
\bea
\av {J_\vx^i}=\int {\cal D}\xi \, P_{\rm prop.}(\xi)P_{\rm accept}(\xi)\, J_\vx^i
\label{eq:met_j_1}
 \ea
where the measure is defined such that $\int {\cal D}\xi P_{\rm prop.}(\xi)=1$. Then noise is gaussian as given in Eq.~\eqref{eq:met_noise}. Therefore the proposal distribution is simply
 \bea
 P_{\rm prop.}(\xi)= \exp\left(-\frac12 \sum_{\vxp,j} (\xi_{\vxp}^{j})^{2}\right),
 \ea
In the continuum limit, the change in the entropy at each Metropolis step is infinitesimally small, so the acceptance probability can be expanded as
 \bea
 P_{\rm accept}(\xi)={\rm min}(1,e^{\Delta S}) \approx \Theta(\Delta S)+\Theta(-\Delta S)\left(1+\Delta S\right)=1+\Theta(-\Delta S)\Delta S\,.
 \label{eq:accept_reject}
 \ea
Given that the proposed current is proportional to $\xi$, the first term in the right hand side of Eq.~\eqref{eq:accept_reject} (i.e. 1) does not contribute the integral in Eq.~\eqref{eq:met_j_1} due to parity and only the second term, $\Theta(-\Delta S)\Delta S$, contributes:
 \bea
\av {J_\vx^i}&=& \sqrt{\frac{2\lambda}{\Delta t \Delta V}} \, \int {\cal D}\xi \, e^{-\frac12 \sum_{\vxp,j} \xi_{\vxp}^{j2}}\, \left[\Theta(-\Delta S)\Delta S \right] \,\xi_{\vx}^i 
\\
&=& \frac12 \sqrt{\frac{2\lambda}{\Delta t \Delta V}} \, \sum_{\vy,k}\int {\cal D}\xi \, e^{-\frac12 \sum_{\vxp,j} \xi_{\vxp}^{j2}}\,  \left[-\frac{\del S}{\del n_\vy} \frac{\sqrt{2\Delta t\lambda/\Delta V}}{a}  \left(\xi_\vy^k-\xi_{\vy-\hat k}^k\right) \right]\,\xi_{\vx}^i\, 
\\
&=& -\lambda  \, \sum_{\vy,k}\int {\cal D}\xi \, e^{-\frac12 \sum_{\vxp,j} \xi_{\vxp}^{j2}}\,  \frac{1}{a \Delta V} \left[\frac{\del S}{\del n_\vy} -\frac{\del S}{\del n_{\vy+\hat k}} \right]
(\xi_\vy^k \xi_{\vx}^i)\, 
\\
&=& -\lambda  \, \sum_{\vy,k}   \frac1a \left[\frac{\del S}{\del n_\vy} -\frac{\del S}{\del n_{\vy+\hat k}} \right] \delta_{\vx\vy}\delta^{ik}= -\lambda  \frac{1}{a \Delta V} \left[\frac{\del S}{\del n_\vx} -\frac{\del S}{\del n_{\vx+\hat i}} \right] 
 \label{eq:met_j}
\ea
The factor of $1/2$ in the second line is due to the fact that only terms with $\Delta S <0$ contribute to the integral. Finally, by using the local thermodynamic relation, $\Delta V^{-1}\del S/\del n=-\mu/T=-\alpha$, in the continuum limit we can write  
\bea
 \frac{1}{a \Delta V} \left[-\frac{\del S}{\del n_{\vx+\hat i}} +\frac{\del S}{\del n_\vx} \right]= \frac1a \left[\alpha_{\vx+\hat i} -\alpha_\vx \right]\rightarrow \frac{\del \alpha}{\del x_i}\,.
\ea
Therefore we see that the resulting the average current takes the correct dissipative form in the continuum limit 
\bea
\av{\bf J}=-\lambda \nabla \alpha \,.
\ea
Furthermore, as a consistency check, a similar calculation shows that the fluctuation dissipation theorem is also obeyed:
\bea
\av{ J^i_\vx  J^j_\vy}=\frac{2\lambda}{\Delta t \Delta V}\delta_{\vx \vy}\delta_{i j}
\ea

There are several advantages of using the Metropolis algorithm over the Langevin equation.  Firstly the infinite noise problem is diluted by the accept-reject step. Even though the magnitude of the noise is still large in comparison to the deterministic part, configurations which involve large changes are very likely rejected. The renormalization of the bare equation of state and transport coefficients still have to be carried out. However, since the infinite noise problem is mitigated, the renormalization procedure is not as challenging to carry out numerically compared to Langevin. Secondly, the fluid variables are sampled in accordance with the correct equilibrium distribution (even for finite $\Delta t$) by construction. This eliminates the discretization ambiguities associated with multiplicative noise from the start. Finally, because there is no dissipative term in the deterministic step, the issues with stability that arise in first order hydrodynamics are effectively bypassed. We further elaborate on this point later.

Recently, the Metropolis algorithm was used to tackle different important problems in stochastic fluid dynamics. It was used to analyze relaxational (model A) \cite{Schaefer:2022bfm} and diffusional (model B) \cite{Chattopadhyay:2023jfm} models of dynamical critical phenomena with non-conserved and conserved order parameters respectively \cite{Hohenberg:1977}. In both cases, the dynamical critical exponents were computed:
\bea
z=2.026(56)\,\, \rm{(model \, A)}, \quad z = 3.906\,\,\rm{(model \,B)}\,.
\label{eq:z_metro}
\ea
These values happen to be in agreement with the $\eps$ expansion. It is worth pointing out that results based on the $\eps$ expansion rely on a truncation of the Feynman diagrams at the one-loop (or two-loop) order. However, higher loops are not parametrically suppressed in the limit $\eps\rightarrow 1$ (i.e. d=3). In other words, the $\eps$ expansion is parametrically uncontrolled in the $\eps\rightarrow 1$ limit,  The Metropolis algorithm, on the other hand, is non-perturbative by construction and there is no uncontrolled expansion involved. Both the infinite volume and continuum extrapolations can be carried out in a controlled fashion. From this perspective it is remarkable that the results Eq.~\eqref{eq:z_metro} are in agreement with the $\eps$ expansion. At the same time, it is consistent with the observation that $\eps$ expansion captures the experimentally observed phenomena fairly well \cite{Hohenberg:1977}. 

In addition to the dynamical critical exponents, equilibration rates for higher moments of the order parameter were studied in model A by quenching the system \cite{Schaefer:2022bfm}. This is achieved by setting the system in thermal equilibrium at high temperature and instantaneously switching the temperature to the critical value. It was observed that the relaxation rate for different moments was different. Furthermore in model B, Kibble-Zurek type scaling was observed by a different quench procedure, namely, by reducing the temperature with a fixed rate as opposed to switching it instantaneously \cite{Chattopadhyay:2023jfm}. These studies were generalized to model H, which in addition to relaxation dynamics exhibits fluid dynamics as well \cite{Chattopadhyay:2024jlh}. Due to the existence of the fluid modes, Model H has rich dynamics that involve coupling between diffusive and shear modes, known as ``mode coupling".  Historically, studies of mode coupling has relied on uncontrolled truncations of the loop expansions similar to models A and B \cite{Hohenberg:1977}. In \cite{Chattopadhyay:2024jlh} the renormalization of the viscosity as well as the dependence of the dynamical critical exponent, $z$, on the viscosity was studied via the Metropolis approach. It was seen that irrespective of how small the the bare viscosity is, the renormalized viscosity in fact has a minimum value consistent with the prediction of the one-loop result \cite{Kovtun:2011np,Chafin} where the fluctuation correction to the bare viscosity, $\eta_0$, scales as $1/\eta_0$ as sketched in Eq.~\eqref{eq:renormalization}. 
Furthermore, it was found that the dynamical critical exponent varies between $3$ and $4$ depending on the value of the viscosity.

Another avenue where Metropolis algorithm was used is chiral dynamics in the context of relativistic hydrodynamics \cite{Florio:2021jlx,Florio:2023kmy}. The main idea stems from the observation that in the limit where the light (up/down) quark masses are zero, QCD exhibits a continuous chiral phase transition in the $O(4)$ universality class \cite{pisarski-wilczek,Rajagopal:1992qz}. In the real world, where light quark masses are not zero but small, it might be possible to observe the signatures of this transition in the correlations in soft pion distribution in heavy ion collisions \cite{Grossi:2021gqi}, since the $O(4)$ order parameter is identified with the pion field. The dynamical critical phenomena associated with the chiral transition is identified to be in the universality class of model G \cite{Rajagopal:1992qz} and a hydrodynamical picture of the pion field was developed in \cite{Son:2001ff}. The dynamics of $O(4)$ have been studied in detail in \cite{Florio:2021jlx,Florio:2023kmy}. The system possesses non-trivial dynamics due to the non abelian nature of the conserved charges which are associated with the generators of $O(4)$ as well as their coupling to the order parameter, similar to the mode coupling in model H. From the real-time dynamics realized via Metropolis algorithm, the value of the dynamical critical exponent has been computed and was found to be consistent with the model G prediction $z=d/2$ for $d=3$ \cite{Hohenberg:1977}. Furthermore, a new set of dynamical critical ratios that relate the diffusion constant to a suitably defined order parameter relaxation time have been determined \cite{Florio:2023kmy}. 

For relativistic fluids, the Metropolis algorithm presents a conceptual challenge, which is the absence of a global time. For an arbitrary fluid flow defined by some time-like vector $u$, it is not always possible define a foliation of space-time such that each Metropolis update is evaluated on a given space-like hyper-surface orthogonal to $u$. A solution to this issue has was presented in \cite{Basar:2024qxd}. The idea is to formulate the problem in a fixed frame. A convenient choice the so called ``density frame" introduced in \cite{Armas:2020mpr} to describe fluids without boost symmetry. The advantage of the density frame is that the dissipative terms have no time-like components, and as a result of the fluctuation-dissipation theorem, the noise is accordingly purely spatial in this frame. This property eliminates the second order derivative terms in the equations of motion which lead to instabilities and makes it possible to simulate fluctuating hydrodynamics with a general flow at first order in derivative expansion. The disadvantage is that the system is explicitly Lorentz non-covariant. However, violation of Lorentz covariance appears in high frequency modes which lie outside of the hydrodynamic regime anyway. 

The first step in this program was taken in \cite{Basar:2024qxd} where nonlinear diffusion in the presence of a background flow was studied.  This system has advection in addition to diffusion. The current in the density frame is defined such that its time component is identified with the charge density, $n_{DF}$, namely 
\bea
 \sJ_{DF}^\mu=(\sn_{DF},  \sn_{DF}v^i- \slambda^{ij}\nabla_j \salpha+\sI^i)\,.
\ea
Here $\bf v$ is the fluid velocity, the first, second and third terms in the spatial current respectively correspond to advection, diffusion with the diffusion matrix $\lambda^{ij}=\lambda(\delta^{ij}-v^iv^j)$, and noise. The noise accordingly is purely spatial with a local, gaussian correlation whose magnitude is fixed by the fluctuation dissipation theorem:
\bea
\av{\sI^i(x)\sI^j(y)}=2\lambda^{ij} \delta^{(4)}(x-y)
\ea 
In this formalism, there is no issue with foliation of space-time and the Metropolis updates can be done exactly in the same way described earlier, with the dissipative tensor being a function of fluid velocity. In contrast, in the Landau frame both the current and therefore the noise has time-like components:
\bea
\sJ_{Lan}^\mu=\sn u^\mu- \slambda \Delta^{\mu\nu}\nabla_\nu  \salpha+I_{Lan}^\mu,\quad \av{\sI^\mu_{Lan}(x) \sI_{Lan}^\nu(y)}=2\lambda \Delta^{\mu\nu}  \delta^{(4)}(x-y)
\ea    
which lead to the aforementioned instabilities. In \cite{Basar:2024qxd}  the effects of the violation Lorentz covariance has been studied by comparing the Metropolis algorithm with a microscopic kinetic model and it was found that, as expected, these effects occur in the high frequency sector of the model. Remarkably, it was shown that the density frame predictions are quite reasonable even in the extreme case where the system size is comparable to the mean free path where one expects the UV modes to be significant. Furthermore, how the inclusion of higher order gradient terms  improves the regime of validity of hydrodynamics, and eventually restores causality upon resummation was explicitly demonstrated. This was done by studying the analytical form of the \textit{all-orders} gradient expansion in the density frame.  This mathematical analysis supports the observation that even the first order truncation has a remarkable practical utility. The extension of these ideas to full hydrodynamics is a promising future direction, currently being pursued.

%% file: hydrokinetic.tex
\subsection{General formulation}
An alternative approach to simulating directly the stochastic equations of fluctuating hydrodynamics is to describe the fluctuations via a set of \textit{deterministic} evolution equations.
These equations constitute an extension of ordinary hydrodynamics by extra modes that characterize fluctuations. The extra modes are not conserved quantities, such as energy-momentum or particle number, but they are correlation functions of them. Their dynamical evolution describes the equilibration of fluctuations in an evolving fluid background. 
One remarkable advantage of this framework is that the infinite noise problem and renormalization of the equation of state and transport coefficients can both be dealt with analytically, before numerically evaluating the equations. The fluctuation evolution equations were first derived for Bjorken flow in Refs. \cite{Akamatsu:2017,Martinez:2017fbb,Akamatsu:2018,Martinez:2018}. They were  generalized to accommodate an arbitrary fluid background in Refs. \cite{An:2019rhf,An:2019fdc,An:2020jjk}, and non-Gaussian fluctuations in Refs. \cite{An:2020vri,An:2022jgc}.  

Before we go into the details of the deterministic framework let us first outline the main idea. To keep the notation simple, we will denote the hydrodynamic fields (pressure, specific entropy and fluid velocity), collectively as $\phi$ as before. We start from the stochastic form of the conserved currents given in the previous section:
\bea
 \sT^{\mu\nu}&=&\sth\eps\sth u^\mu\sth u^\nu+\sth p\sth\Delta^{\mu\nu} +\sth \Pi^{\mu\nu} +\sS^{\mu\nu} \nn
  \sJ^{\mu}&=&\sth n\sth u^\mu+ +\sJ_D^{\mu} +\sI^{\mu\nu}
  \ea
Classical hydrodynamics simply describes the evolution of the average values of the currents, $\av{\sT^{\mu\nu}}$ and $\av{\sJ^\mu}$. Provided that the fluctuations are small in the hydrodynamic limit (we will later quantify the smallness) we can expand these expressions around the average values of the primary variables:
\bea
\sth \phi= \phi+\dphi\,.
\ea 
Here $\phi:=\av{\sth \phi}$ denotes the average value and $\dphi$ is the deviation from it, i.e. the fluctuation. Furthermore, because the equation of state and the constitutive relations depend non-linearly on $\sth\phi$, in general we have 
\bea
\av{\sT^{\mu\nu}}\neq T^{\mu\nu}[\av{\sth \phi}],
\ea
and the average currents includes contributions of higher order correlation functions:
\bea
\av{\sT^{\mu\nu}}&=& T^{\mu\nu}[\phi(x)]+\frac{\delta^2 T^{\mu\nu}}{\dphi_{a_1}\dphi_{a_2}} \av{\dphi_{a_1}(x)\dphi_{a_2}(x)}+\frac{\delta^3 T^{\mu\nu}}{\dphi_{a_1}\dphi_{a_2}\dphi_{a_3}} \av{\dphi_{a_1}(x)\dphi_{a_2}(x)\dphi_{a_3}(x)}+\dots \\ 
\label{eq:Tav}
\av{\sJ^{\mu}}&=& J^{\mu}[\phi(x)]+\frac{\delta^2  J^{\mu}}{\dphi_{a_1}\dphi_{a_2}} \av{\dphi_{a_1}(x)\dphi_{a_2}(x)}+\frac{\delta^3  J^{\mu}}{\dphi_{a_1}\dphi_{a_2}\dphi_{a_3}} \av{\dphi_{a_1}(x)\dphi_{a_2}(x)\dphi_{a_3}(x)}+\dots
\label{eq:Jav}
\ea
The locality of the correlators comes from the fact that the constitutive relations are local. Which means that the contribution from the higher order correlation functions is singular as they diverge when evaluated at the coincident points.. The leading divergence comes from the equilibrium contribution as discussed is in Sec. \ref{sec:equilibrium}.In equilibrium, the two-point function behaves as
 \bea
 \av{\dphi_{a_1}(x)\dphi_{a_2}(x)}^{\rm eq}\sim\delta^{3}(0)\sim \Lambda^3\,.
 \ea
 Here $\Lambda^{-1}$ is the inverse size of the hydrodynamic cell and plays the role of the short-distance (i.e. UV) cutoff.  This short-distance singularity is in fact nothing but the manifestation of the infinite noise problem. Here the divergent contributions can be calculated explicitly and furthermore be absorbed in the appropriate terms in the constitutive relations. This way, in the deterministic approach, renormalization can be achieved analytically. When the evolution equations expressed in terms of renormalized quantities, they contain no divergences which is a significant advantage when numerically simulating them.

In addition to the singular terms, there are also finite, cutoff-independent contributions from the higher point functions to the average stress energy tensor. They are non-local in space-time (alternatively they have the form $\omega^{3/2}, k^{3/2}$ in Fourier space), and constitute the leading corrections to the viscous terms in the gradient expansion (which are of order $k$). These non-local fluctuation induced terms encode the famous long-time tails \cite{Kovtun:2003,Kovtun:2011np}. We will come back to this point after discussing the evolution equations for the two-point functions.  

The main objects of interest in the deterministic approach are the \textit{equal-time} correlation functions which measures the correlation between the hydrodynamic modes different points at equal time:
\bea
G_{a_1a_2}(t,\vx_1,\vx_2)&:=&\av{\dphi_{a_1}(t,\vx_1)\dphi_{a_2}(t,\vx_2)} \\
  G_{a_1a_2 a_3}(t,\vx_1,\vx_2,\vx_3)&:=&\av{\dphi_{a_1}(t,\vx_1)\dphi_{a_2}(t,\vx_2)\dphi_{a_3}(t,\vx_3)},\\
  &\vdots& \nonumber
\ea 
For now we will focus on the two-point correlation functions. The higher-point functions are discussed in Sec. \ref{sec:nongaussian}. 
It is useful express the fluctuations in Fourier space. This can be achieved by a Wigner transform where only the relative separation between the two points (or the ``fast variable''), is Fourier transformed. This fast variable characterizes the fluctuations. Meanwhile, the average of the two points (the ``slow variable"), which characterizes the background flow, remains in position space.  Therefore the Wigner transform is defined as
\bea
W_{a_1a_2}(t,\vx;\vq)=\int d^3 \vy \,e^{- i\vq\cdot\vy} \,\av{\phi_{a_1}(t,\vx+\vy/2) \phi_{a_2}(t,\vx-\vy/2)}\,.
\label{eq:two_point_wigner}
\ea
Here  we used a mixed index $a_i$ to denote all the fluctuation modes. When $a_i$ takes the value $m$ or $p$ it behaves like a scalar index, wheres when it denotes the components of the fluctuations of fluid-velocity, $a_i=\mu$, it behaves like a vector index. The Wigner transform was generalized to higher-point correlators in Ref. \cite{An:2020vri}. One useful property of the Wigner function is that in equilibrium, it takes a $\vq$ independent value since the equilibrium fluctuations are local. This property generalizes to the higher-point functions as well as discussed in Sec. \ref{sec:nongaussian}.

The fluctuation evolution equations describe how the Wigner function evolves in time. As we discuss below, their form resembles the Boltzmann equation for in kinetic theory where  $\vx$ and $\vq$ play the role of the phase space coordinates. Of course, these ``particles" are not the microscopic constituents of the fluid but phonons that are associated with fluctuation modes. For this reason, these evolution equations are dubbed as ``hydro-kinetic" equations \cite{Akamatsu:2017}. The hydro-kinetic evolution equations for the two-point functions in relativistic hydrodynamics were first derived in the pioneering work of Ref.  \cite{Akamatsu:2017} for Bjorken flow. In Ref. \cite{Martinez:2017fbb} they were generalized to Bjorken plasma  with conserved charge density.  In the context of non-relativistic hydrodynamics they were studied in \cite{Andreev:1978}. 

 In the background of a generic fluid flow the evolution equations are quite complicated \cite{An:2019rhf,An:2020jjk}. However, remarkably, they can still be put into a form of kinetic-like equations where various components of the two-point function after appropriate rescaling can be identified with the distribution of phonons associated with fluctuations. The spectrum of phonons is parameterized by a four vector which can be decomposed into energy and wave-vector of the phonon in the local rest frame of the fluid. We can explicitly write this four-vector as $-Eu^\mu+q^\mu$. Here $q^\mu$ represents the wave-vector obtained via the Wigner transform satisfying $u\cdot q=0$, and $E:=c_s|q|$ represents the phonon energy.
 
 In general the two-point function $W_{ab}$ has many components. Some of these components rapidly oscillate with frequencies $\omega \sim c_s |q|$ and at time scales where the fluctuations evolve and reach equilibrium, these modes average out to zero. The modes that remain have longer lifetimes and have the typical frequency dependence $\omega \sim -i \gamma_d |q|^2$.
  There are seven such slow modes and they are further decomposed into one sound mode and six diffusive modes. The kinetic equation for the sound mode is found to be \cite{An:2020jjk}
  \bea
\left[(u+v)\cdot \cfd+f\cdot\pd{}{q}\right] W_{\rm sound}=-\gL q^2 \left[W_{\rm sound}-W^{\rm eq}_{\rm sound}\right]+K W_{\rm sound}
\label{eq:Wsnd}
\ea
where the phonon velocity is given as
\bea
v^\mu=\pd{E}{q_\mu}=c_s\hat q^\mu\,.
\ea
 and $K$ contains terms proportional to fluid gradients. The left hand side of Eq.~\eqref{eq:Wsnd} takes the form of a Liouville operator with the force term given as  
 \bea
 f^\mu=-E(a^\mu+2\omega^{\mu \nu} v_\nu)-q^\nu \del_\perp^\mu u_\nu-\cfd_\perp^\mu E\,.
 \label{eq:liouvillef}
 \ea
 Here $a_\mu$ is the acceleration, and $\omega^{\mu\nu}$ is the vorticity, leading to the relativistic generalization of the Coriolis force. The third term  describes the deformation of the sound wave due to the expansion. It can be understood by considering an isotropic, uniform expansion, $\del_\perp^\mu u^\nu=H \Delta^{\mu\nu}$ with $H$ being the expansion rate. This third term then becomes
 \bea
 q^\nu \del_\perp^\mu u_\nu \pd{}{q^\mu}= H q \cdot\pd{}{q}
 \ea
 which is analogous to the photon redshift in the expanding universe. 
 
 Also, notice the bar above the gradient operator in Eqs. \eqref{eq:Wsnd} and \eqref{eq:liouvillef}. It denotes a differential operator called ``confluent derivative". This new kind of derivative is defined in Sec. \ref{sec:confluent}. For now it suffices to say that it is necessary to describe how fluctuations vary over space-time in a Lorentz covariant way. The right-hand side describes the relaxation to equilibrium $W^{\rm eq}_{\rm sound}=Tw$. As expected, each mode relaxes with a rate proportional to the magnitude of the wave-vector, $q^2$, the larger wavelength modes equilibrating slower. 

In addition to the sound mode discussed here, the six diffusive modes are proportional to $W_{mm}$, $W_{m(i)}$ and $W_{(i)(j)}$ where $(i)$ denotes the directions orthogonal to both $u$ and $q$. Each of these modes have zero phonon velocity and relax with rates $\gL q^2$, $(\geta+\gL) q^2$ and $\geta q^2$ respectively, namely,
\bea
{\cal L}[W_{mm}] &=&-\gL q^2 \left[W_{\rm sound}-c_p T^2\right]+ \dots\\
{\cal L}[W_{m(i)}]&=&-(\geta+\gL) q^2  W_{m(i)}+\dots\\
{\cal L}[W_{(i)(j)}]&=&-\gL q^2 \left[W_{(i)(j)}-Tw \delta_{ij} \right]+\dots\\
\ea 
where $\dots$ denote the background terms linear in gradients of the fluid variables. They are lengthy terms that couple $W_{mm}$, $W_{m(i)}$ and $W_{(i)(j)}$ nontrivially. The exact expressions for them can be found in Section 3 of \cite{An:2019fdc}.

\subsection{Bjorken expansion}

The first calculations of the fluctuation evolution equations and their feedback on the background (long-time tails) were done in the Bjorken background. Bjorken flow describes a fluid expanding in a boost invariant way in one direction, and is homogeneous and isotropic in the plane transverse to the expansion.  The space-time dynamics is best expressed in the Milne coordinates,
\bea
x^\mu=(\tau, \vx_\perp,\eta)
\ea
where $\tau=\sqrt{(x^0)^2-(x^3)^2}$ is the proper time, and $\eta={\rm arctanh}(x^3/x^0)$ is the rapidity\footnote{Note the unfortunate but standard notation for the rapidity which shares the same letter as the shear viscosity. }. The metric is given by $g_{\mu\nu}=(-1,1,1,\tau^2)$ with non-vanishing Cristoffel symbols $\Gamma^\tau_{\eta\eta}=\tau$, $\Gamma^{\eta}_{\eta\tau}=\Gamma^{\eta}_{\tau\eta}=\tau^{-1}$.  The fluid expands in the $x^3$ direction. The boost invariance condition means that all the dynamical quantities depend only on the proper time $\tau$.  The hydrodynamic limit implies that the gradient expansion can be trusted in the limit of large $\tau T$, where the temperature, $T$, is determined as a function of $\tau$ by the ideal hydrodynamic equations. As the fluid expands it cools down. The energy density, for instance, decreases in time by 
\bea
\epsilon(\tau)=\epsilon_0 \left(\frac{\tau_0}{\tau}\right)^{1+c_s^2}\left(1+ \dots \right)
\ea
where the dots denote the viscous corrections which scale as $1/\tau$. 

Out of the 12 components of a general symmetric two-point function of conserved energy and momentum, only four of them are relevant. The rest either vanish due to symmetry or rapidly oscillate and average out to zero in the typical timescale of fluctuations as mentioned earlier. The kinetic equation for these modes are best expressed as longitudinal (denoted by $\pm$) and transverse (denoted by $(i)$) components.\footnote{In the original references, \cite{Akamatsu:2017,Akamatsu:2018}, the transverse modes are denoted by $T_{i}$ with $\in\{1,2\} $ rather than $(i)$.}. The kinetic equations are found as \cite{Akamatsu:2018}
\bea
\pd{N_{\pm\pm}}{\tau}&=&-\gL \vq^2\left(N_{\pm\pm}-\frac{Tw}{\tau}\right) -\frac1\tau\left(2+c_s^2+T\frac{dc_s}{dT}+\cos^2\theta\right) N_{\pm\pm} \\
\pd{N_{(1)(1)}}{\tau}&=&-2\geta \vq^2\left(N_{(1)(1)}-\frac{Tw}{\tau}\right) -\frac2\tau N_{(1)(1)} \\
\pd{N_{(2)(2)}}{\tau}&=&-2\geta \vq^2\left(N_{(2)(2)}-\frac{Tw}{\tau}\right) -\frac2\tau(1+\sin^2\theta) N_{(2)(2)}
\label{eq:bjorken_Ns}
\ea
where $\theta$ denotes the polar angle.
Here the two point function $N_{ab}$ is proportional to the Wigner transform of the two-point function as above. Also there is no conserved charged in this setup, and the longitudinal diffusion constant is  $\gL=(4\eta/3+\zeta)/w$. The first terms in each equation describe the relaxation towards equilibrium and the second terms are due to the expansion of the fluid. Note that due to the expansion, the equilibrium value itself is time dependent, $Tw\sim\tau^{-(1+2c_s^2)}$, highlighting the dynamical nature of the problem. Notably the left hand side in the evolution equations, the Liouville operator, simply reduces to $\del_\tau$. The aforementioned ``Hubble term" in the Liouville operator that governs the redshift due to the expansion in this case is $q^3/\tau \del_{q_3}$ since  the expansion rate for Bjorken flow is $\del_\mu u^\mu=1/\tau$. This term can be eliminated by rescaling $q_3\rightarrow \tau q_3$, in other words working with the Fourier transform of rapidity instead of $x^3$.     

The contribution of the fluctuations to the energy density and pressure follow from Eq.~\eqref{eq:Tav}. For simplicity, focusing on a conformal fluid where $c_s^2=1/3$ and $\zeta=0$, the contributions of fluctuations to energy density and longitudinal pressure are found as \cite{Akamatsu:2017}
\bea
\av{\sT^{\tau\tau}}=\epsilon+\frac{1}{w}\av{\dvg^2},\quad 
\av{\sT^{\tau\tau}}=c_s^2\epsilon-\frac{4\eta}{3\tau}+\frac{1}{w}\av{\dg_z^2},
\ea
where $\dvg=(T^{\tau x},T^{\tau y},\tau T^{\tau \eta})$ denotes the momentum fluctuations. In the Bjorken background the average momentum flow vanishes due to isotropy; $\av{\sth {\bf g}}=0$. These contributions can be calculated by solving the evolution equations and then integrating them over the phase space. As explained in the previous section the short-distance singularities lead to cubic and linearly divergent terms which can be absorbed into the renormalization of the equation of state and shear viscosity.

After the singular terms are absorbed, the remaining finite part of $N_{ab}$ peaks around  $q_*\sim1/\sqrt{\tau \geta}$.  Denoting the finite term as $\wt N_{ab}$, upon integrating over phase space,we have
\bea
\int d^3q \wt N_{ab} \sim q_*^3 \sim \frac{1}{(\tau \geta)^{3/2}}
\ea
Note that these terms are higher order than the viscous terms which scale as $1/\tau$ but lower order than the second order terms in the hydrodynamic gradient expansion $\sim 1/\tau^2$. The exact results for the long time tails for Bjorken flow are found as  \cite{Akamatsu:2017}
\bea
\tau^2 \av{\sth T^{\eta\eta}}&=&p-\frac{4\geta}{3\tau}w+T \frac{1.08318}{\left(4\pi\geta\tau\right)^{3/2}} +\dots 
\\
 \av{\sth T^{xx}}= \av{ \sth T^{yy}}&=&p+\frac{2\geta}{3\tau}w-T\frac{0.273836}{\left(4\pi\geta\tau\right)^{3/2}}+\dots
\\
 \av{\sth T^{\tau\tau}}&=& \av{\sth T^{xx}}+\av{ \sth T^{yy}}+\tau^2 \av{ \sth T^{\eta\eta}}+\dots
\label{eq:bjorken_ltt_t}
\ea
where the dots refer to the contribution form second order terms that are ${\cal O}(1/\tau^{2})$. Here the viscosities and the equation of state are renormalized.

The same formalism was extended to Bjorken flow with a conserved charge \cite{Martinez:2018}. In this case, in addition to the energy density and pressure, there are also fluctuation induced contributions to the charge current. The time-like component gives the correction to the charge density, whereas the space-like components give the corrections to the Ohm current induced by an external electric field. For the Bjorken flow, they are given as \cite{Martinez:2018}
\bea
 J^\tau&=&n+\frac{Tn}{w}\frac{0.04808}{\left(\geta\tau\right)^{3/2}}
\\
\frac{ J^x}{E^x}=\frac{ J^y}{E^y}&=&\sigma-\frac{Tn^2\tau}{w^2}\left[\frac{0.0266}{\left(\geta\tau\right)^{3/2}}+\frac{Tc_p}{w}\frac{0.03558}{\left((D+\geta)\tau\right)^{3/2}}\right]
\\
\frac{\tau  J^\eta}{\tau E^\eta}&=&\sigma-\frac{Tn^2\tau}{w^2}\left[\frac{0.04282}{\left(\geta\tau\right)^{3/2}}+\frac{Tc_p}{w}\frac{0.008}{\left((D+\geta)\tau\right)^{3/2}}\right]
\label{eq:bjorken_ltt_j}
\ea
where ${\bf E}$  denotes the external electric field,  $\sigma=\lambda/T$ is the conductivity and $D=\sigma[(\del \mu/\del n)_\eps -\alpha (\del T/\del n)_\eps]$. 

\subsection{General background flow}
\label{sec:general_flow}

The general fluctuation evolution equations for the two-point function in the presence of arbitrary background flow were derived in Refs. \cite{An:2019rhf,An:2019fdc}. The generalization is far from straightforward and requires a set of new mathematical tools to describe the fluctuations in a Lorentz covariant way which is dubbed as the ``confluent formalism".  Here we highlight to main results for the two-point functions which were obtained using the confluent formalism. The details of this framework are discussed in Sec. \ref{sec:confluent}.

As mentioned earlier, the fluctuation evolution equations can be neatly recast as Liouville equations for propagating sound modes and diffusive modes. The back-reaction of the fluctuations on the hydrodynamic gradient expansion follows from the solutions of these equations evaluated at the coincident limit (see Eq.~\eqref{eq:Tav}). Following this procedure, in Ref.  \cite{An:2019fdc}, the fluctuation contribution to the energy momentum tensor and the current in general background was calculated as 
\bea
\label{eq:piomega}
\av{\sth \Pi^{\mu\nu}}&=&\Pi^{\mu\nu}+\frac{1}{w} {\wt G}^{\mu\nu}(x)+\frac{\bD^{\mu\nu}}{2}\left(\frac{(1-\dot c_p)}{c_pT}\wt G_{mm}(x)+\frac{c_s^2-\dot T+2\dot c_s}{w}\wt G_{pp}(x)-\frac{c_s^2+\dot T}{w}{\wt G}^\lambda_\lambda(x)\right)\qquad
\\
\av{\sth J_D^{\mu}}&=&J_D^\mu-\frac{n}{w^2}\wt G^{m\mu}(x)-\frac{c_s n}{w^2}\wt G^{p\mu}(x)\,.
\ea
where $\Pi^{\mu\nu}$ and $J_D^\mu$ are the classical dissipative terms given in Eqs. \eqref{eq:dissipative_Pi} and \eqref{eq:dissipative_J}. Here $\wt\,$ denotes the finite part of the correlator after the cubic and linearly divergent terms are absorbed into the renormalized equation of state and viscosities respectively.  We also used the notation $\dot X:=(\del \log X/\del \log s)_m$ for brevity. 

Recall that the finite contribution to the correlator comes from subtracting the cubic and linear short-distance divergences. In $\vq$ space, these terms respectively behave as constant (equilibrium contribution) and  $1/q^2$. The non-equilibrium contribution is obtained after subtracting the $q$ independent equilibrium contribution. From Eq.~\eqref{eq:Wsnd} one can deduce that the non-equilibrium part schematically can be written as
\begin{equation}
  \label{eq:Wneq}
    W^{\rm neq}\equiv W-W^{\rm eq} \sim 
\frac{\partial f}{\gamma q^2+i( u + v)\cdot  k+\del f}\,
\end{equation}
where $v=\pm c_s\hat q$ or $0$ depending on whether we are considering the sound mode or a diffusive mode. $\gamma$ and $\del f$ respectively represent the generic relaxation rate and terms linear in background gradients.  Moreover, $k \sim \del$ and $u\cdot k=\omega$ represent the typical wave-vector associated with the background flow and the typical frequency.  The linearly divergent part is proportional to the gradients of background and is absorbed into the renormalization of transport coefficients. Subtracting it from the non-equilibrium contribution leads to 
\begin{equation}
  \label{eq:Wtilde}
   \wt W \sim 
\frac{\partial f}{\gamma q^2+i( u + v)\cdot  k+\del f}-\frac{\partial f}{\gamma q^2}\,
\end{equation}
 Finally evaluating the phase space integral yields the non-analytic behavior in the gradient expansion
\bea
\wt G \sim \int d^3 q \wt W\sim \frac{k^{1/2} \del f}{\gamma^{3/2}} \sim \frac{k^{3/2}}{\gamma^{3/2}}
\ea
The long time tails are encoded in the in the zero wavelength and finite frequency limit. By using these steps that are outlined above, in Ref. \cite{An:2019fdc} the non-analytic contribution analytic contributions to the frequency dependent conductivity and viscosity were found as 
\bea
\lambda(\omega)&=&\,\lambda -\omega^{1/2}\frac{T^2n^2}{w^2}\frac{(1-i)}{6\sqrt{2}\pi}\left(\frac{c_pT}{(\geta+\gl)^{3/2}w} +\frac{c_s^2}{2\gL^{3/2}}\right)
\\
\eta(\omega)&=&\,\eta-\omega^{1/2}T\frac{(1-i)}{60\sqrt{2}\pi}\left(\frac{1}{\gL^{3/2}}+\frac{7}{(2\geta)^{3/2}}\right)
\\
\zeta(\omega)&=&\,\zeta-\omega^{1/2}T\frac{(1-i)}{36\sqrt{2}\pi}\left(\frac{1}{\gL^{3/2}}\left(1-3\dot T+3\dot c_s\right)^2+\frac{4}{(2\geta)^{3/2}}\left(1-\frac{3}{2}(\dot T+c_s^2) \right)^2+\frac{9}{2(2\gl)^{3/2}}\left(1-\dot c_p\right)^2\right).\nn
\label{eq:viscosities_omega}
\ea
These expressions characterize the long-time tails in the presence of an arbitrary fluid background and equation of state. In the particular limit of Bjorken flow and conformal equation of state they reduce to the expressions given the previous section.

\subsection{Multi-point Wigner function}
As explained earlier, the characteristic scale of fluctuations is shorter than that of the hydrodynamic background. For the two-point function, the background, described by the average coordinate $\vx=(\vx_1+\vx_2)/2$, evolves slower than the fluctuations, described by the relative coordinate $\vx_1-\vx_2$ which is Fourier transformed via a Wigner transformation given in Eq.~\eqref{eq:two_point_wigner}. In this section we discuss how this formalism generalizes to the higher-point correlators in a natural way which was introduced in Ref. \cite{An:2020vri}.
 
   A generic equal-time, $n$-point correlator, 
 \bea
 G_\adotsn(t,\vx_1,\dots \vx_n)=\av{\phi_{a_1}(t,\vx_1)\dots\phi_{a_n}(t,\vx_n)},
 \ea
  depends on $n$ coordinates $\vx_i$. The dependence on the mid-point $\vx=(\vx_1+\dots +\vx_n)/n$ occurs in the longer, background scale. The remaining $n-1$ coordinates, hence $n-1$ independent wave-vectors, characterize fluctuations. At the same time, it is convenient keep the expressions in a symmetric form where we introduce a set of $n$ wave-vectors $(\vq_1,\dots \vq_n)$ and define the Wigner transform as
\bea
W_\adotsn(t,\vx;\vq_1,\dots ,\vq_n)=\int \prod_{i=1}^n \left[ d^3 \vy_i \,e^{- i\vq_i\cdot\vy_i} \right]\delta^{(3)}((\vy_1+\dots\vy_n)/n)\av{\phi_{a_1}(t,\vx+\vy_1) \dots \phi_{a_n}(t,\vx+\vy_n) }\,.
\label{eq:Wn_def}
\ea
Here the delta function ensures that the fluctuation coordinates, $\vy_i$s, are defined relative to the mid-point $\vx=(\vx_1+\dots+\vx_n)/n$. It is also due to the same delta function that $W_{a_1,\dots a_n}$ is invariant under a shift of all the $\vq_i$s by the same amount. In other words $W$ does not depend on $\vq_1+\dots +\vq_n$, and it suffices to know $W$ on the $n-1$ dimensional surface $\vq_1+\dots +\vq_n=0$. The inverse Wigner transform is defined as
\bea
G_\adotsn(t;\vx_1,\dots ,\vx_n)=\int \prod_{i=1}^n \left[ \frac{d^3 \vq_i}{(2\pi)^3} \,e^{i\vq_i\cdot \vx_i} \right](2\pi)^3\delta^{(3)}(\vq_1+\dots+\vq_n)W_{a_1\dots a_n}(t,\vx;\vq_1,\dots ,\vq_n) 
\label{eq:Wn_inv_def}
\ea
The partial derivative transforms under the Wigner transform as
\bea
\nabla_{\vx_i} G_n\rightarrow \left(i\vq_i+\frac1n \nabla_{\vx} \right) W_n
\ea

\subsection{Confluent Formalism}
\label{sec:confluent}
Bjorken expansion belongs to a special class of flow profiles where there is a well defined global time parameter, $\tau$, and the fluid velocity can be expressed as a gradient $u^\mu=\del^\mu \tau$. This property allows one to foliate space-time such that for each value of $\tau$ there is a space-like surface which can be identified as the equal-time surface. In general, however, this is not the case. When there is vorticity, it is not possible to foliate space-time in such a way; therefore the notion of the equal-time correlator has to be defined more carefully.

As discussed earlier, the fluctuations in a fluid occur locally in shorter distances compared to the scale of the variation of the background flow.  It is therefore natural to describe them in the local in the rest frame of the fluid. However the notion of the rest frame depends on the spacetime coordinate. In other words, the ``time" direction is associated with the four-velocity of the fluid which varies from point to point. Moreover, the fluid velocity, hence the direction of time, also fluctuates locally. Therefore, the notion of equal-time used in the definition of the Wigner function introduced in Eq.~\eqref{eq:Wn_def} is not well defined for a relativistic fluid with nontrivial gradients. The idea behind the confluent formalism is to unambiguously define the fluctuations in a Lorentz covariant way. It was first introduced in  \cite{An:2019rhf} and was generalized to non-Gaussian fluctuations in \cite{An:2020vri,An:2022jgc}.

The main issue is to define the notion of equal-time for an $n$-point correlator. It is convenient to work in the local rest frame with respect to the \textit{average} fluid velocity, evaluated at the mid-point of the correlator, $x=(x_1+\dots+x_n)/n$. This choice fixes the time direction to be $u(x)$. The equal-time hyper-surface is then spanned by a set of three local basis vectors which we call $\ve(x)$. Each basis vector is a four vector perpendicular to $u(x)$, i.e.
\bea
\ve(x)\cdot u(x)={\bf 0}\,.
\ea
\begin{figure}[h]
\center
\includegraphics[scale=0.5]{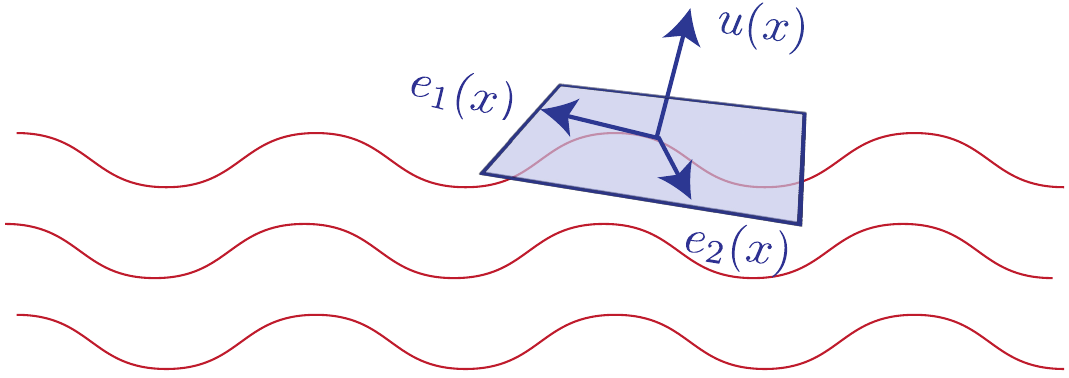}
\caption{At each point $x$ the local time direction is determined by the average fluid velocity $u(x)$ and the spatial hyper-surface is spanned by the local triad $\ve(x)$.}
\label{fig:basis}
\end{figure}
Together with $u$, they form a local basis for the four-dimensional Minkowski space as illustrated in Fig. \ref{fig:basis} We use the index convention introduced in \cite{stephanov2024qcd} to label each basis vector as $e_{\oa}(x)$ with  $\oa=1,2,3$. Since each $e_\oa$ is a four-vector, its components are expressed as $e_\oa^\mu(x)$.  
The coordinates of fluctuations around the mid-point are expressed by a set of spatial vectors, $\vy_i$,
 \bea
 x_i=x+\ve(x)\cdot \vy_i\,.
 \ea
As a result we can define the equal-time correlator to be
\bea
G_\adotsn(x; \vydotsn):=\av{\phi_{a_1}(x_1) \dots \phi_{a_n}(x_n) }
\label{def:Gn_def}
\ea
  Just like in the previous section, the fluctuation coordinates obey the constraint $\vy_1+\dots+\vy_n=0$. 
  
\begin{figure}[h]
\center
\includegraphics[scale=0.5]{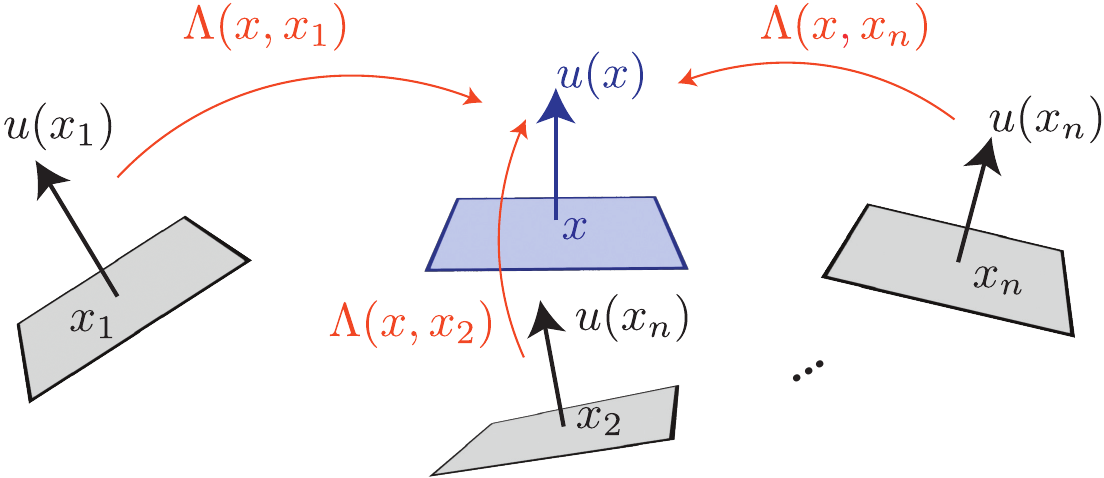}
\caption{An illustration of the confluent correlator. All the vector valued fields are boosted to the common frame defined by the average velocity $u(x)$.}
\label{fig:conf_corr}
\end{figure}

 Recall that $\phi_{a_i}$ is an umbrella term the set of independent, primary hydrodynamic fields. This set includes the fluctuation of velocity $\delta u_\mu (x_i)$ which is a Lorentz vector. Since we have chosen to work in the frame defined by $u(x)$, it is convenient boost these vector valued fields to this frame. To do so, we define  the operator that boosts $u(x_i)$ to $u(x)$ as 
 \bea
 \Lambda(x,x_i)u(x_i)=u(x)
 \ea
 Boosting the vector valued fields in the $n$-point correlator into the common frame defined by $u(x)$ leads to the notion of the \textit{confluent correlator} defined as
 \bea
\bar G_\adotsn(x; \vydotsn):=\av{\left[\Lambda(x,x_1)\phi(x_1)\right]_{a_1} \dots \left[\Lambda(x,x_1)\phi(x_n)\right]_{a_n} }
\label{eq:Gn_conf_def}
 \ea
In order to distinguish it from the ``raw" correlator defined in Eq.~\eqref{def:Gn_def} we denote the confluent correlator with a bar. In Eq.~\eqref{eq:Gn_conf_def} and for the rest of this review, the boost operator acting on scalar components of $\phi$ such as specific entropy or pressure is simply equal to multiplication by unity. The kinematical structure of the confluent correlator is illustrated in Fig. \ref{fig:conf_corr}. We can now define equal-time Wigner function in the covariant form, namely the \textit{confluent Wigner function}, as 
\bea
 W_{a_1\dots a_n}(x;\vq_1,\dots ,\vq_n):=\int \prod_{i=1}^n \left[ d^3 \vy_i \,e^{- i\vq_i\cdot\vy_i} \right]\delta^{(3)}((\vy_1+\dots\vy_n)/n)\bar G_\adotsn(x; \vydotsn)\,.
\label{eq:Wn_conf_def}
\ea
 The inverse of the confluent Wigner function is 
\bea
\bar G_\adotsn(x;\vy_1,\dots ,\vy_n)=\int \prod_{i=1}^n \left[ \frac{d^3 \vq_i}{(2\pi)^3} \,e^{i\vq_i\cdot \vx_i} \right](2\pi)^3\delta^{(3)}(\vq_1+\dots+\vq_n) W_{a_1\dots a_n}(x;\vq_1,\dots ,\vq_n) 
\label{eq:Wn_conf_inv_def}
\ea
 
\begin{figure}[h]
\center
\includegraphics[scale=0.49]{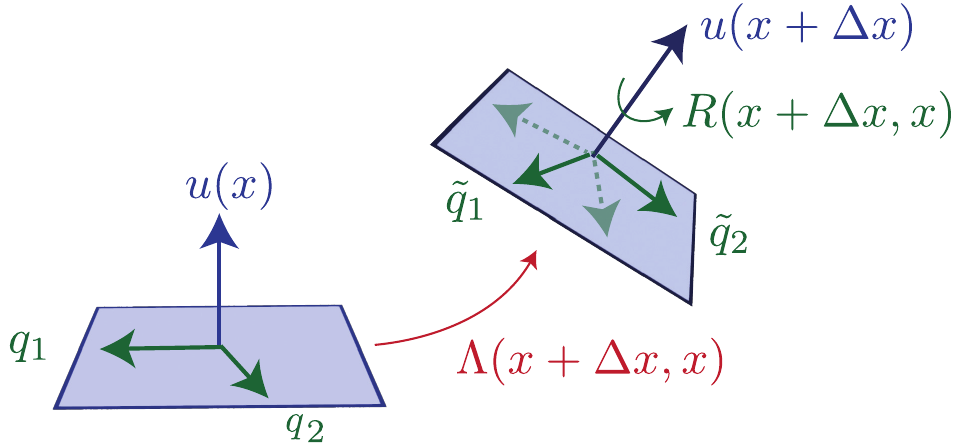}
\qquad
\includegraphics[scale=0.49]{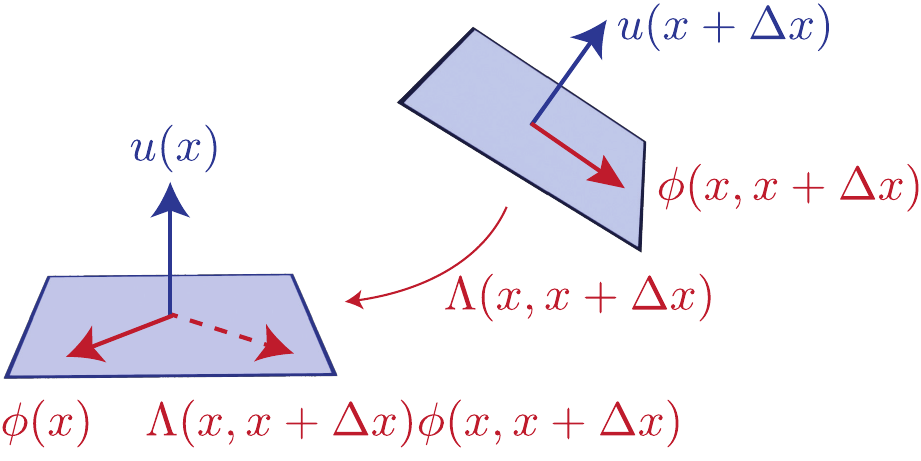}
\caption{Pictorial representations of two ingredients of the confluent connection defined in Eq.~\eqref{eq:conf_connection}: the  ``circle connection", $\econ_{\mu\ob}^\oa$ ,  and the ``boost connection" $\ucon^{b_1}_{\mu a_1}$.  }
\label{fig:omega_connections}
\end{figure}

 The evolution equations for the correlators involve space-time gradients and therefore relate $W_n$ evaluated at different space-time points. 
 At the same time, the notion of ``equal-time" is different at different points since $u$ is a function of $x$. It is therefore useful to define a type of derivative which maintains the equal-time condition when the function varies between $x$ and $x+\Dx$. This can be achieved by comparing with $W_n(x)$, not $W_n(x+\Dx)$ as an ordinary derivative would do, but $W_n(x+\Dx)$ boosted to the local rest frame defined by $u(x+\Dx)$. The operator which does this is called as the \textit{confluent derivative} and denoted by the symbol $\cfd$. It is defined as 
 \bea
 \Dx^\mu\cfd_\mu W_\adotsn(x;\vqdotsn)&=& \Lambda(x,x+\Dx)^{b_1}_{a_1}\dots \Lambda(x,x+\Dx)^{b_n}_{a_n} W_{b_1\dots b_n}(x+\Dx;\tilde\vq_1,\dots,\tilde\vq_n)
 \nn
 &&-W_\adotsn(x;\vqdotsn)
 \label{eq:conf_der_def_discrete}
 \ea 
 with $\Dx^\mu\rightarrow 0$. Notice that the fluctuation wave-vectors also change upon varying $x$. This is because $\vq_i$s are defined on the surface perpendicular to $u(x)$ and in order for the derivative to be meaningfully defined, they have to be parallel transported to the surface perpendicular to $u(x+\Dx)$. This parallel transport can be done by expressing $\vq_i$ as a four-vector via $q_i=\ve(x)\cdot \vq_i$ and boosting it to the new frame by
 \bea \tilde q_i =\Lambda(x+\Dx,x)q_i\,.
 \ea
 Finally, since the wave-vectors associated with $W_n(x+\Dx)$ are perpendicular to $u(x+\Dx)$, we complete the parallel transport by projecting $\tilde q_i$ on the new equal time surface at $x+\Dx$ as $\tilde\vq_i=\ve(x+\Dx)\cdot \tilde q_i$. This operation in general corresponds to an SO(3) rotation:
 \bea
 \tilde \vq_i= \ve(x+\Dx)\cdot \tilde q_i =  \ve(x+\Dx)\cdot  \left[ \Lambda(x+\Dx,x) \ve(x) \cdot \vq_i\right]\
\equiv R(x+\Dx,x)\vq_i
 \ea 
An infinitesimal boost can be written as
\bea
\Lambda^\mu_\nu(x+\Dx,x)=\delta^\mu_\nu-\Dx^\lambda \ucon_{\lambda\nu}^\mu,\quad \ucon_{\lambda\nu}^\mu=u_\nu\del_\lambda u^\mu-u^\mu\del_\lambda u_\nu\,.
\label{eq:ucon}
\ea
Note that, in principle, the boost is defined up to a rotation in the direction of $u$. In Eq.~\eqref{eq:ucon} we made the simplest choice where the boost is not accompanied by any rotation. With this choice, the rotation operator can similarly be expanded to linear order in $\Dx$ as
\bea
R^\oa_\ob(x+\Dx,x)=\delta^\oa_\ob-\Dx^\lambda \econ_{\lambda\ob}^\oa,\quad \econ_{\lambda\ob}^\oa=e^\oa_\mu\del_\lambda e^\mu_\ob
\label{eq:circ}
\ea
In sum, transporting $W_n$ from $x$ to $x+\Dx$ by maintaining its confluent structure can be achieved by two types of connections: the boost connection, $\ucon_{\lambda\nu}^\mu$ defined in Eq.~\eqref{eq:ucon} which boosts  the Lorentz valued fields  (denoted by Greek letters), and the ``circle connection" $\econ_{\mu\oa}^\ob $, which rotates the vectors that live on the spatial hyper-surface perpendicular to $u$  (denoted by circled latin letters, e.g. $\oa$). Therefore, the confluent derivative acting on objects with Lorentz and transverse indices can be written in differential form as
\bea
\cfd_\mu v^\nu_\oa=\del_\mu v^\nu_\oa+\ucon_{\mu\lambda} v^\lambda_\oa-\econ_{\mu\oa}^\ob v^\nu_\ob\,.
\ea
In particular, the confluent derivative of the multi-point Wigner function is given as
\bea
\cfd_\mu W_\adotsn(x;\vqdotsn)=\del_\mu W_\adotsn(x;\vqdotsn)
- n\left(\ucon^{b_1}_{\mu a_1}W_{b_1\ldots a_n}-\econ^\oa_{\mu\ob} q_{1\oa}\frac{\partial}{\partial q_{1\ob}}W_\adotsn \right)_{\apermn}\,.
\,\label{eq:conf_connection}
\ea
 Here the subscript denotes the sum over all $n!$ permutation of the composite indices $(a_1,x_1),\dots,(a_n,x_n)$ divided by $n!$,
 \bea
 \big(\dots\big)_\apermn=\frac{1}{n!} \big(\dots\big)_{{\rm P}a_1\ldots a_n},
 \ea 
 and $\del_\mu$ refers to the ordinary partial derivative that keeps $\vq_i$s constant, namely $\Dx^\mu \del_\mu W_n(x,\vqdotsn)=W_n(x+\Dx,\vqdotsn)-W_n(x,\vqdotsn)$ with $\Dx^\mu\rightarrow 0$.  In Eq.~\eqref{eq:conf_connection} the boost connection boosts the velocity fluctuation into the new equal time frame and the circle connection takes into account the rotation of the fluctuation wave-vectors upon the parallel transport as illustrated in Fig.~\ref{fig:omega_connections}. 
 
 Notably, by definition, the average fluid velocity and the three basis vectors orthogonal to it are constant with respect to the confluent connection
 \bea
 \cfd_\mu u^\nu=0, \quad \cfd_\mu  e^\nu_\oa=\del_\mu e^\nu_\oa +\ucon_{\mu\lambda}^\nu e^\lambda_\ob -\econ_{\mu\oa}^\ob e^\nu_\ob=0.
 \label{eq:du_de_confluent}
 \ea
 One can, alternatively, impose Eq.~\eqref{eq:du_de_confluent} as a starting point and define the connections accordingly. This way one arrives, up to a rotation for the boost connection as we discussed, at Eqs. \ref{eq:ucon} and \ref{eq:circ}. 

 In short, the confluent connection can be thought of an operation that measures the change of the fluctuating variables with respect to the local rest frame that they measured in and not the change in their components that occur due to the change in the local rest frame.

 \subsection{Non-gaussian fluctuations}
 \label{sec:nongaussian}

The equations for higher-point correlation functions in the presence of background flow using the machinery developed in the previous two sections were derived in \cite{An:2020vri,An:2022jgc}. The first reference focuses on the non-linear diffusion problem and the second one focuses on the fluctuations of the specific entropy, the slowest mode in the vicinity of the critical point, including the effects fluctuations of pressure.  

As an example, the two-point function of the specific entropy satisfies the relaxation equation 
\bea
\label{eq:W_mm}
    \mathcal{L}[W_{mm}(\vq)]=\theta W_{mm}(\vq)-2\gamma_{mm} \vq^2\left[ W_{mm}(\vq)-\frac{c_p}{n} \right]\,,
\ea
where the Liouville operator takes the form
\bea
{\cal L}[W_{mm}(\vq)]:=\left[u\cdot\cfd+{\bf f}(x,\vq)\cdot\pd{}{\vq} \right] W_{mm}(\vq)
\label{eq:liuoville_wmm}
\ea
with $\bar\nabla$ being the confluent derivative (see Eqs. \eqref{eq:conf_der_def_discrete} and \eqref{eq:conf_connection}). Here since $W_{mm}$ has no Lorentz indices, the boost connection, $\ucon$, does not enter the expression for the confluent derivative. The force term captures the kinematical effects due to the fluid background such as expansion or shear flow and is given as 
\bea
{\bf f}(x,\vq)=- (\ve_\mu\cdot\vq) \ve^\nu \del_\nu u^\mu\,.
\ea
Finally, the appropriate diffusion constant for this mode is identified as
\bea
\gamma_{mm}:=\gamma_\lambda=\frac{\kappa}{c_p}
\ea
Notice that near the critical point $c_p\sim \xi^2$ diverges, so that the equilibrium value of  $W_{mm}$ diverges. At the same time, the diffusion constant goes to zero, making this mode slowest to reach equilibrium, highlighting the importance of its dynamics. The equations for the three and four-point functions of the specific entropy have been derived in \cite{An:2022jgc}. 

In order to discuss the general form the equation for the higher point functions and highlight the key steps in how they are obtained, it is more transparent to consider the nonlinear diffusion problem:   
\bea
\del_t \sn=-  \nabla \cdot  ({\bf  \sJ}_D+ {\bf   \sI})\,.
\label{eq:sth_diff}
\ea
The dissipative current in the lab frame is simply ${\bf J}_D=-\lambda \nabla \alpha$. In general, the diffusion constant $\lambda$ and $\alpha=\mu/T$ depend non-linearly on the conserved density $n$. The noise satisfies 
\bea
\av{\sI_i(x_1) \sI_j(x_2)}=2\lambda \delta_{ij}\delta^{(4)}(x_1-x_2)
\ea
As usual, defining the deviation from the average value as
\bea
\dn=\sn-\av{\sn}
\ea
we can write the stochastic diffusion equation, Eq.~\eqref{eq:sth_diff}, compactly as 
\bea
\label{eq:dtn}
  \del_t \dn= F[\dn]+\xi \,,
  \quad\mbox{with}\quad
  \av{\xi(x_1)\xi(x_2)}=Q\delta^{(4)}(x_1-x_2)\,.
\ea
Here the drift term 
\bea
  F:= -  \nabla \cdot  {\bf  \sJ}_D+\av{\nabla \cdot  {\bf  \sJ}_D}
 \ea 
 is in general a nonlinear (but local) functional of $\dn$. The noise term, $Q$, is a second order differential operator which follows from 
 \bea
 \xi:=-\nabla\cdot {\bf \sI}\,.
 \label{eq:xi}
\ea
As mentioned in Sec. \ref{sec:stc_general}, the nonlinearities of fluctuations in the hydrodynamic regime are suppressed by the central limit theorem. Therefore it is useful to expand the drift term in fluctuations  
\bea
  F=\sum_{k=1}^\infty   F_k [ \dn^k]\,.
\ea 
Here the $F_k$s are multilinear, second-order differential operators acting on $\dn^k$ whose coefficients are expressed as combinations of the derivatives of the diffusion constant $\gamma:=\lambda \alpha^\prime$ and conductivity $\lambda$. This form of the Langevin equation Eq.~\eqref{eq:dtn} generalizes in a straightforward way to the case where there are multiple conserved quantities \cite{An:2022jgc}. 

The characteristic magnitude of fluctuations is inversely proportional to $1/\sqrt{N}$ where $N$ is the number of uncorrelated cells. Away from any critical region where fluctuations are enhanced, the typical correlation length is at the order of the microscopic length scale, $\lmic$. Therefore typical fluctuations are suppressed by the factor 
\bea
\eps:=\frac{1}{\sqrt N } \sim \frac{1}{\sqrt{{\ell}_{\rm f}/\lmic^3}} \sim (q\lmic)^{3/2}
\ea
where $q$ is the magnitude of the typical fluctuation wave-vector. In addition to the smallness of the magnitude of fluctuations controlled by $\varepsilon$, the gradients of fluctuations are also small in the hydrodynamic limit. The gradient expansion is controlled by another small parameter which we call
\bea
\eps_q:=q\lmic\,.
\ea
 Of course, in the hydrodynamic limit $\eps$ and $\eps_q$ are related, but it is useful to treat them separately because there are examples where these two quantities are parametrically different. For example in the large $N_c$ limit of QCD, the fluctuations are suppressed by $1/\sqrt{N_c}$. At leading order in $\eps$, a generic $n-$point function
and a generic \textit{connected} $n-$point function (denoted with the superscript c) scale as  
 
\bea
  G_{n}\sim\varepsilon^{[n/2]}\,,\quad
  G^c_{n}\sim\varepsilon^{n-1}
  \label{eq:Gn_epsilon}
  \ea
  where $[n/2]=n/2$ for even $n$ and $(n+1)/2$ for odd $n$. The drift and noise terms similarly scale as 
  \bea
  F_k  \sim \varepsilon_q^2, \quad
  Q \sim\varepsilon_q^2\varepsilon\,\,.
  \ea
The quadratic dependence on $\eps_q$ is a consequence of the conservation law: for the drift term one gradient comes from the conservation Eq.~\eqref{eq:sth_diff} and another one from the constitutive equation. For the noise term it comes from Eq.~\eqref{eq:xi}. In full hydrodynamics the contribution of the advective part of the constitutive relation to $F_k$ is ${\cal O}(\eps_q)$ while the dissipative and noise terms are still ${\cal O}(\eps_q^2)$.

\begin{figure}[h]
\center
\includegraphics[scale=0.5]{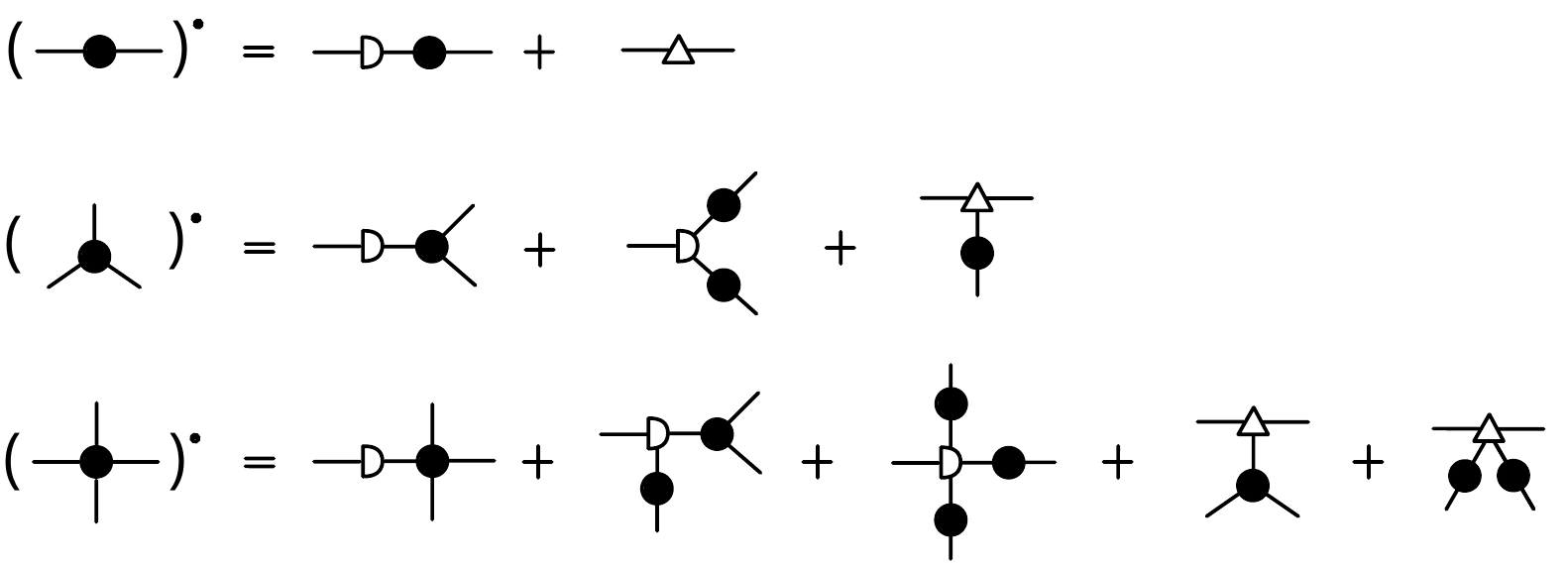}
\caption{ The diagrammatic representations of the evolution equations for the 2,3, and 4 point functions to leading order in fluctuations. }
\label{fig:diagram}
\end{figure}

\begin{figure}[h]
\center
\includegraphics[scale=0.5]{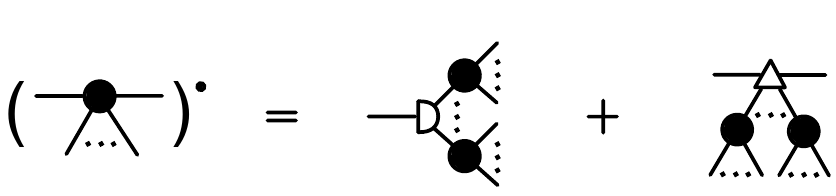}
\caption{ The diagrammatic representation of the fluctuation evolution equations for the n-point correlator to leading order in fluctuations. The left-hand side denote the time derivative (or Liouville operator in the presence of background fluid flow) and the dots on the right hand side denote all possible tree-level combinations that has $n$ external legs. 
}
\label{fig:diagram_gen}
\end{figure}

\begin{figure}[h]
\center
\includegraphics[scale=0.5]{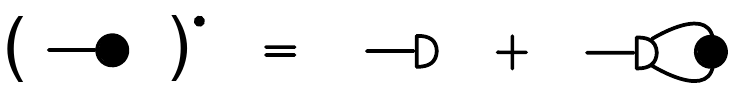}
\caption{ The evolution of the one point function. The first term represents the ordinary hydrodynamic flow and the second term, the loop correction, represents the effect of fluctuations and encodes long time tails. The second term is suppressed by a factor of $\eps$ compared to the first one. 
}
\label{fig:loop}
\end{figure}

The evolution equation for the equal-time $n-$point function follows from evaluating the time derivative $\del_t \av{\dn(t,\vx_1)\dots \dn(t,\vx_n)}$ by using the Langevin equation given in Eq.~\eqref{eq:dtn} for each $\dn$. Whether one uses It$\bar {\rm o}$  or Stratonovich calculus in this procedure does not matter, as long as the drift term is properly adjusted such that the stationary solutions reproduce the correct equilibrium distribution fixed by thermodynamics \cite{Arnold:1999va}. Using the It$\bar {\rm o}$ prescription we have
\bea
\del_t \av{\dn(t,\vx_1)\dots \dn(t,\vx_n)}&=& \av{ ( F[\dn]+\xi)_{\vx_1} \dn(t,\vx_2)\dots\dn(t,\vx_n)}+\dots + \av{ \dn(t,\vx_1)\dots( F[\dn]+\xi)_{\vx_n} }\nn
&&+  \av{ \xi({\vx_1}) \xi(\vx_2) \dn(t,\vx_3)\dots\dn(t,\vx_n)} +\dots
\label{eq:Wn_equation_general}
\ea
Note that, in general, because $\delta F$ depends on $\dn$ non-linearly, \textit{all} higher point functions appear on the right-hand side of Eq.~\eqref{eq:Wn_equation_general}. However, in the hydrodynamic limit the contribution of higher point functions are suppressed by extra powers of $\eps$. Therefore, to leading order in $\eps$, the equation for a given $n-$point function contains only $n$ and lower point functions. Furthermore, loop corrections are also higher order in $\eps$ and can be neglected as well. As a result, the evolution equation for $G_n$ contain all possible combinations of the form $G_{k_1}\times \dots \times G_{k_m}$ which are of order $\eps^{[n/2]}$. More explicitly, the equations for the Wigner transformed two, three, and four point functions are \cite{An:2020vri}
\bea
\label{eq:W_234-b}
 \del_tW_2(\vq_1,\vq_2)\
       &=&-2\left[\gamma\vq_1^2W_2(\vq_1,\vq_2)+\lambda \vq_1\cdot \vq_2\right]
              \,
       \\
      \del_tW_3(\vq_1,\vq_2,\vq_3)\
       &=&-3\left[\gamma\vq_1^2W_3(\vq_1,\vq_2,\vq_3)
         +\gamma'\vq_1^2W_2(-\vq_2,\vq_2)W_2(-\vq_3,\vq_3) \right. \nn
        && \left. +2\lambda'\vq_1\cdot\vq_2W_2(-\vq_3,\vq_3)\right]_{\overline{123}}
       \,
       \\
        \del_tW_{4}(\vq_1,\vq_2,\vq_3,\vq_4)
       &=&
       -4\left[
         \gamma\vq_1^2W_{4}(\vq_1,\vq_2,\vq_3,\vq_4)  +3\gamma'\vq_1^2W_2(-\vq_2,\vq_2)W_3(-\vq_3-\vq_4,\vq_3,\vq_4)
             \right.
         \nn
         &&  \left.
    +\gamma''\vq_1^2W_2(-\vq_2,\vq_2)W_2(-\vq_3,\vq_3)W_2(-\vq_4,\vq_4)
    +3\lambda'\vq_1\cdot\vq_2W_3(-\vq_3-\vq_4,\vq_3,\vq_4)  \right.
    \nn
  && 
  \left.   +3\lambda''\vq_1\cdot\vq_2W_2(-\vq_3,\vq_3)W_2(-\vq_4,\vq_4)
   \right]_{\overline{1234}}
    \ea
To keep our notation compact, we suppressed the dependence on $t$ and $\vx$ (the average coordinate). Also the Wigner transform is defined such that the fluctuation wave vectors sum to zero, i.e. $\vq_1+\dots+\vq_n=0$.  These equations have a simple diagrammatic representation as shown in Fig. \ref{fig:diagram}. Here the solid circle with $k$ legs represents $W_k$. The semi-circle with $k+1$ legs and the triangle with $k+2$ legs respectively represent $F_k$ and $\delta^{k} Q/\dn^{k}$. In other words, the semicircle terms encode the contribution of drift and the triangle terms encode the contribution of the noise, including the effects of multiplicative noise which manifests itself in the equations for three and higher point functions as the triangle with $k>2$ legs.
Generalizing the form we obtained for the two, three, and four point functions, we conclude that the equation for a generic $n-$point function has two ingredients, drift and noise. Each of these ingredients can be represented by a sub-class of diagrams where all possible tree-level combinations of $n-$ and lower point functions appear provided that that the total number of legs add up to $n$. This is illustrated in Fig. \ref{fig:diagram_gen}. Remarkably, the hierarchical structure where only $n-$ and lower point functions appear on the right-hand side of the equation for the $n-$point function, which is a consequence of the hydrodynamic limit, allows these equations be solved iteratively. The equation for $W_2$ can be solved on its own since it only depends on $W_2$. Then its solution is fed into the equation for $W_3$ as an inhomogeneous source term. After solving for $W_3$, now $W_3$ and $W_2$ become source terms for the equation for $W_4$ and so on. This iterative structure makes the deterministic approach of studying fluctuations practically appealing.
As mentioned above, in the diagrammatic language, the feedback of fluctuations, which encode the long time tails, appear as loop diagrams. Compared to the leading tree-level diagrams, they are suppressed by further powers of $\eps$. For example the effect of the two-point function on the background hydrodynamic flow, which corresponds to the second term in Eq.~\eqref{eq:Tav}, is represented diagramatically as show in Fig. \ref{fig:loop}.

Coming back to the fluctuations of specific entropy, the equations for $W_{mm}=\av{\delta m\delta m}$ can be directly mapped to the nonlinear diffusion problem by identifying $W_2 \rightarrow nW_{mm}$, $\gamma\rightarrow \gamma_{mm}$ and $\lambda\rightarrow\kappa/n$. For the higher point functions, the equations for $n^2W_{mmm}$ and $n^3W_{mmmm}$ are qualitatively similar to those of $W_3$ and $W_4$ in the nonlinear diffusion problem\footnote{The factors $n$, $n^2$ and $n^3$ in the map between $W_{m\dots m}$ and $W_n$ allows one to eliminate the kinematic contributions $\theta W_{mm}$, $2\theta W_{mmm}$, $3\theta W_{mmmm}$  in the evolution equations for $W_{m\dots m}$, such as the one in Eq.~\eqref{eq:W_mm}.}, however unlike $W_{mm}$, this map only captures the terms which are leading in the scaling regime. In general, there are other terms which cannot be reproduced by mapping to the nonlinear diffusion problem due to the nonlinear relation between the specific entropy and the conserved density $n$. These terms were calculated in Ref. \cite{An:2022jgc}

The evolution equations of non-gaussian correlators resemble the equations for the cumulant that have been studied in Refs. \cite{Berdnikov:1999ph,Mukherjee:2015swa}. Those cumulants correspond to $W_n$s that are integrated over the phase space, whereas the evolution equations in Eq.~\eqref{eq:W_234-b} characterize the relaxation of a continuous family of modes parameterized by the set of wave-vectors $\vq_1,\dots ,\vq_n$. It can be clearly seen from the equations that the longer wavelength modes relax slower than the shorter wavelength modes. Likewise, it can also be seen that higher point functions also relax faster. Furthermore, three and higher point functions are characterized by multiple wave-vectors and depending on their relative magnitudes and angles, there is a quite rich dynamical structure. In general, assuming an isotropic system, an $n-$point function can be described by $n(n-1)/2$ independent invariants, $\vq_i\cdot \vq_j$, with $i<j$ that parameterize the phase space. 

\subsection{Critical fluctuations in hydro-kinetics}
\label{sec:critical_hydrokinetics}

Near a critical point, the correlation length, $\xi$ diverges and various thermodynamic quantities, such as $c_p$ diverges. This behavior affects both the magnitude of fluctuations, as well as relaxation rates. Furthermore, not every mode grows or slows down equally as the thermodynamic quantities that are associated with their magnitudes and relaxation rates scale differently with $\xi$. The slowest and largest mode corresponds to the fluctuations of specific entropy as mentioned. We discussed stochastic simulations of various models near criticality in Sec. \ref{sec:stochastic}. Here we outline some approaches of implementing critical dynamics in the hydro-kinetic framework.  An alternative approach to hydro-kinetic analysis of the specific entropy fluctuations is reviewed in the next section. 

In order to accommodate the effects of the critical phenomena hydro-kinetic equations have to be modified. Firstly the rate relaxation rate, $\Gamma=\gamma_d q^2$ in Eq.\eqref{eq:W_234-b},  has to reflect the critical slowing down. In Ref. \cite{Akamatsu:2018tcp} an ansatz inspired by model B was used
 \bea
\Gamma=\frac{\kappa}{c_p} (1+(q\xi)^{2-\eta_I}) q^2
\ea
 Here $\kappa_0$ is the thermal conductivity and $\eta_I\approx 0.036$ is the correlation length critical exponent of 3d Ising model. In Ref. \cite{An:2019fdc} a similar ansatz based no model H was used
 \bea
 \Gamma=\left(\frac{\kappa}{c_p} (1+(q\xi)^{2})+ \frac{T}{6\pi \eta \xi} K(q\xi) \right)q^2
 \ea
where $K(x)=3/(4x^2)\left[1+x^2+(x^3+x^{-1})\arctan x\right]$ is the Kawasaki function which originates from a one-loop treatment of the mode coupling in model H. Here $\eta_I$ is approximated to zero. In both these anstaze, the $q$ dependent heat capacity appears:
\bea
c_p(q):=\frac{c_p}{1+(q\xi)^{2-\eta_I}}
\ea 
where in the latter it takes the ``Ornstein-Zernike" form with $\eta_I=0$. This modification takes into account the long-range correlations that the equilibrium fluctuations exhibit near a critical point:
\bea
W_{mm}^{eq}:= c_p(q) 
\ea
With these modifications for the equilibrium distribution and relaxation rate, in Ref. \cite{An:2019fdc}, the dynamical contribution of fluctuations to the viscosities and conductivity has been calculated in the same fashion as renormalization of transport coefficients. Furthermore, by including not only the slowest mode $W_{mm}$, but the other modes, $W_{m(i)}$ and $W_{(i)(j)}$ as well, with similar modifications, a model for fluid with a critical point was introduced in the same reference. This model provides a description of the dynamics for frequencies $\omega \lesssim \xi^{-1}$. In contrast ordinary hydrodynamics breaks down for frequencies $\omega \sim\xi^{-3}$ because the relaxation rate of the slowest mode that drives the system out-of-equilibrium at this scale is $\Gamma \sim\gamma_d q^2\sim \xi^{-1}\times \xi^{-2}$ where the first factor comes from the divergence of the heat capacity. Including the dynamics of the slowest modes pushes the applicability to frequencies $\omega \lesssim \xi^{-2}$ where the next slowest modes, $W_{m(i)}$ and $W_{(i)(j)}$ , fall out of equilibrium. Inclusion of these modes pushed the applicability of hydro-kinetics to frequencies $\omega \lesssim \xi^{-1}$. 

In Ref. \cite{Akamatsu:2018tcp} the effects of the interplay between the critical slowing down and the expansion of the fluid was studied. On the one hand the critical fluctuations slow down and grow in magnitude as one approaches a critical point. On the other hand if the fluid expands rapidly as in the heavy ion collisions, or the early universe, it passes near the critical point rapidly and these fluctuations do not get a chance to grow indefinitely. The characteristic time scale of this competition is given by the ``Kibble-Zurek time" where the expansion rate and relaxation rate of fluctuations are comparable, 
\bea
\frac{\del_t c_p}{c_p} \sim \Gamma (t,q)
\ea 
In this timescale, the correlation length can only grow up to certain size, $\xi_{KZ}$. Here $\Gamma$ depends on time as well since the thermodynamic variables that enter $\gamma_d$ vary in time due to expansion. In Ref. \cite{Akamatsu:2018tcp}  by estimating the typical value of $\xi_{KZ}$ it was found that in heavy ion collisions one expects the growth in two-point function in the critical region to be of factor 2 or so. 

\begin{figure}[h]
\center
\includegraphics[scale=0.5]{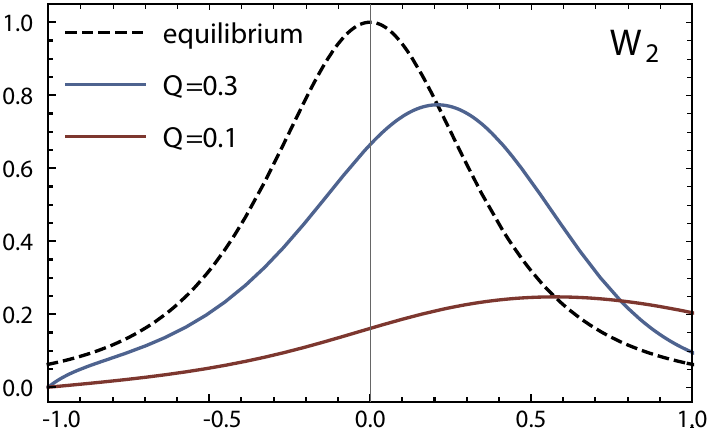}
\includegraphics[scale=0.5]{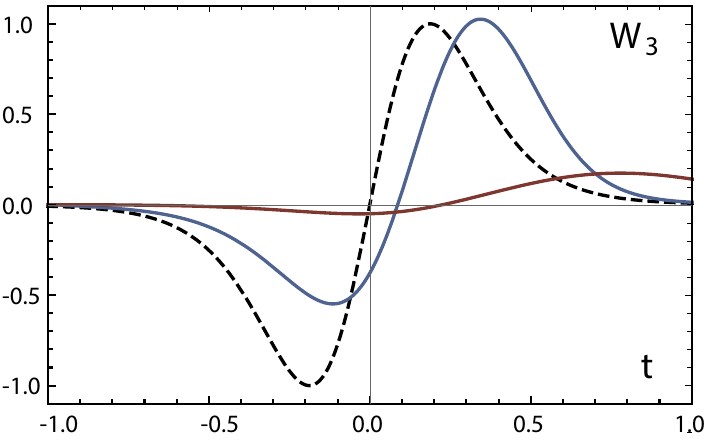}
\includegraphics[scale=0.5]{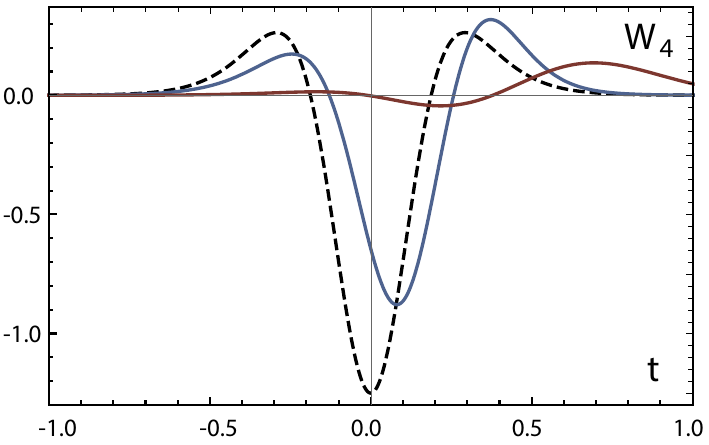}
\caption{  The evolution of $W_{2,3,4}$ in compared with their equilibrium values (dashed line) for an expanding system. The deviation from equilibrium and the time lag becomes more prominent for soft (low $q$) modes. Figure from \cite{An:2020vri}
}
\label{fig:wplots}
\end{figure}

For non-gaussian fluctuations, a simple yet instructive example was studied in Ref. \cite{An:2020vri} to illustrate how the dynamical effects depend on the typical fluctuation scale. In this example, all of wave-vector invariants that characterize an $n-$point function are taken to be equal, i.e. $\vq_i\cdot\vq_j=Q^2$. The dynamics background is chosen to mimic a trajectory of an expanding system that passes near a critical point at $t=0$ where the equilibrium fluctuations peak.  The solutions to the fluctuations evolution equations, with similar modifications to accommodate critical dynamics described above, for this expanding system are plotted in Fig. \ref{fig:wplots}. One can see that the non-equilibrium modes show a memory effect where they reach their peak values later than the equilibrium fluctuations. Furthermore, the memory effect becomes more significant for longer wavelength modes. This is a consequence of the conservation law such that the longer wavelength modes relax slower. It is also notable that the strength of the signal depends on the fluctuation scale. Both these properties are important to consider for observing fluctuations in for example heavy ion collisions.

\subsection{Hydro +}
Following the philosophy of hydro-kinetics, a formalism of incorporating an out-of-equilibrium mode to hydrodynamics (hence the name Hydro +) was developed in Ref. \cite{Stephanov:2018hydro+}.
Hydro + approach is centered around the notion of non-equilibrium entropy density, $s_{(+)}$ which describes a system that is only partially equilibrated. 
In this system, there is a set of slow variables $\varphi_q$, indexed by the parameter $q$, whose values differ from their equilibrium values $\varphi^{\rm eq}_q$.

 The fluid variables such as pressure and viscosities in general depend on these out-of-equilibrium modes and they take different values than they do when $\varphi_q$ is in equilibrium. 
 To emphasize this, we denote the $\varphi_q$ dependent values with the same label, e.g. $p_{(+)}$, $\eta_{(+)}$, $\zeta_{(+)}$. The effect of non-equilibrium modes in hydrodynamics is then captured by 
 the ordinary hydrodynamic equations where the constitutive relations are expressed in terms of the modified fluid variables;
\begin{equation}
  \label{eq:constitutive+}
  T_c^{\mu\nu} = \eps u^\mu u^\nu + p_{(+)} \Delta^{\mu\nu} + \Pi_{(+)}^{\mu\nu}\,.
\end{equation}
The first law also is modified to reflect the existence of these modes via
\begin{equation}
  \label{eq:ds+}
  ds_{(+)}  = \beta_{(+)}d\eps - \alpha_{(+)}dn -  {\bf \pi}\cdot d\varphi\,,
\end{equation}
where we used the compact notation $ {\bf \pi}\cdot \varphi:= \int_q {\bf \pi}_q \varphi_q$. The thermodynamic conjugate of $\varphi_q$ is denoted by $ {\bf \pi}_q$. 
The dynamics of $ \varphi_q$  is described a relaxation equation towards which can be written as, 
\begin{equation}\label{eq:udphi}
   (u\cdot\partial) {\bf \varphi}_q = -F_{\varphi}+ K_\varphi \theta.
 \end{equation} 
Here $K_\varphi$ denotes the background term that is due to the expansion of the fluid. The structure of the drift term, $F$, in this equation is constrained by the second law such that it takes an Onsager form, $(F_\varphi)_q=\int_{q'}\Gamma_{qq'}\pi_{q'}$ with a semi-positive-definite $\gamma$. For systems near equilibrium one intuitively expects $F_q=-\Gamma_q( \varphi_q-\varphi^{\rm eq}_q)$ with some relaxation rate $\Gamma_q$. This is indeed the case as shown in Ref. \cite{Stephanov:2018hydro+}. In this reference the non-equilibrium contribution to the entropy is calculated as 
\begin{equation}
  \label{eq:s+-s}
  s^{\rm neq} := s_{(+)} - s
  =\frac{1}{2}\int_q\left(
    \log\frac{\varphi_q}{\varphi^{\rm eq}_q}
    -\frac{\varphi_q}{\varphi^{\rm eq}_q}+1\right)\,.
     \end{equation}
Remarkably, this expression has a simple interpretation as the \textit{relative entropy} (also known as the Kullback-Liebler divergence in statistics) between the equilibrium and non-equilibrium distributions,
\bea
s^{\rm neq} =-S[{\cal P}|{\cal P}^{eq}]= \sum_{i\in \rm states} {\cal P}_i \log({\cal P}_i/{\cal P}_i^{eq})
\ea
where ${\mathcal P}$ and ${\cal P}^{\rm eq}$ respectively are non-equilibrium and equilibrium distribution functions. From this expression the thermodynamic force follows as 
  \begin{equation}
  \label{eq:pi-phi}
  \pi_q \equiv -\frac{\partial s_{(+)}}{\partial\varphi_q}
   = \frac12\left(\frac1{\varphi^{\rm eq}_q} -
    \frac1{\varphi_q}\right)
  = \frac12{\left(\varphi^{\rm eq}\right)^{-2}}\left( \varphi_q- \varphi^{\rm eq}_q \right)
  + \mathcal O\left(\left( \varphi_q- \varphi^{\rm eq}_q \right)^2\right)\,,
\end{equation}    
 reproducing the expected behavior of the drift term. 
 
 In the vicinity of a critical point the slowest mode is identified with the fluctuations of specific entropy with the identification
 \begin{equation}
  \label{eq:phiN}
  \varphi_\vq(x) = W_{mm}(\vx,\vq)\,. 
\end{equation}
 Here the index $q$ is identified with the fluctuation wave-vector $\vq$. In Ref. \cite{An:2019fdc} it was shown that the Hydro + formalism reproduces the evolution equation of $W_mm$ which was derived from stochastic hydrodynamics. This is a quite nontrivial cross-check of the agreement between two different methods of approaching deterministic analysis of fluctuations. Furthermore it was also shown that the modified fluid variables such as $p_{(+)}$ also correctly reproduces the feedback of fluctuations generated by the fluctuations of specific entropy, given in Sec. \ref{sec:general_flow}. 
 
The equations of Hydro + was numerically studied in a simple but instructive, heavy-ion motivated background with a critical point in Ref. \cite{Rajagopal:2019xwg}. 
This numerical study shows similar non-equilibrium effects to those discussed in Sec. \ref{sec:critical_hydrokinetics}, namely the slower relaxation of longer wavelength modes,
and the memory effect which causes $\varphi_{\vq}$ to lag behind its equilibrium value. 
Moreover advective effects due to the coupling of $\phi_{\vq}$ to the expansion of the fluid are also demonstrated in this reference. Due to advection, the peak of $\varphi_{\vq}$ is observed to move radially outward as one would expect from an expanding plasma.
Moreover, recently Ref. \cite{Abbasi:2021rlp} observed an enhancement of the speed of sound due to the fluctuations by analyzing the spectrum of linear perturbations in Hydro +.

%% file: effective.tex
\subsection{General formulation of Schwinger-Keldysh formalism}
So far we have discussed, what could be described as a ``bottom-up" perspective on the hydrodynamic fluctuations where the hydrodynamic equations of motion are promoted to stochastic evolution equations.  The hydro-kinetic evolution equations were  derived from these stochastic equations. An alternative, ``top-to-bottom" approach is to formulate fluctuating hydrodynamics as an effective field theory (EFT). 
In this section we will review recent progresses along this direction and illustrate how it overlaps with the stochastic as well as hydro kinetic approaches. The idea of formulating relativistic hydrodynamics in terms of an action principle is not new (see for example Ref. \cite{Brown:1992kc}). More recently Refs. \cite{PhysRevD.85.085029,Bhattacharya:2012zx,Banerjee:2012iz,Jensen:2012jh,Haehl:2013hoa} formulated effective actions for ideal relativistic fluids.  At the same time, writing hydrodynamics as a type of statistical field theory is more aligned with our task of studying fluctuating hydrodynamics. Of course, when there are fluctuations, the fluctuation-dissipation theorem necessitates the existence of dissipation. A field theoretical approach to dissipative hydrodynamics have been taken in in Refs. \cite{Endlich:2012vt,Kovtun:2014hpa,Haehl:2014zda,Haehl:2015pja,Crossley:2015evo,Glorioso:2016gsa,Glorioso:2017fpd,Jensen:2017kzi,Haehl:2018lcu,Jain:2020hcu}. A comparison between different approaches that highlights the overlap between these works by different groups can be found in Ref. \cite{Haehl:2017zac}.  A detailed and pedagogical introduction to the EFT approach can be found in Refs. \cite{Kovtun:2012rj, Liu:2018kfw}. Some of these ideas have also been used in improved relativistic formulation of hydrodynamics which preserved causality and stability \cite{Jain:2023obu} as well as non-relativistic systems \cite{Jain:2020vgc}.

Since hydrodynamics is an effective description of a thermal system near-equilibrium, an appropriate medium for the EFT is the Schwinger-Keldysh formalism which originally was developed to formulate non-equilibrium quantum field theory, but can as well be applied to classical stochastic systems  \cite{Kamenev_2011}  such as fluctuating hydrodynamics. Here we outline its basic properties. 

In this framework, fluctuating hydrodynamics is formulated as an effective statistical field theory\footnote{Here we do not consider quantum fluctuations and direct the interested reader to \cite{Liu:2018kfw}.} that describes the slow, long wavelength modes of a thermal system. This field theory emerges after one integrates out the UV degrees of freedom (i.e. fast, high frequency modes) of the underlying microscopic system. In a thermal system, the remaining slow modes describe the conserved quantities (e.g. energy-momentum, particle number) and their fluctuations since all the non-conserved modes thermalize in a shorter time scale as explained earlier in Section \ref{sec:stochastic}. Following the spirit of effective field theory, the goal is to write down the most general action, $I$, consistent with the symmetries of the system, organized as a gradient expansion in the hydrodynamic fields $\Phi$. We will see that there are novel symmetries associated with the thermal nature of the system that are crucial in organizing this expansion. The EFT framework therefore provides a systematic, first-principles approach to the study of the fluctuations. 

\begin{figure}[h]
\center
\includegraphics[scale=0.6]{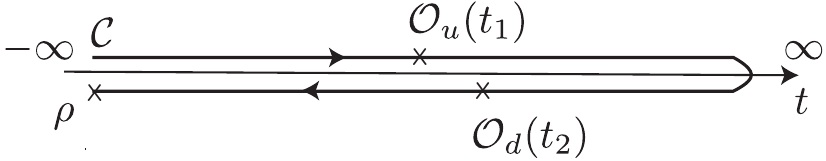}
\caption{A representation of the Schwinger-Keldysh contour with operator insertions at two different times $t_1$ and $t_2$.  $\rho$ represents the insertion of the density operator.}
\label{fig:cpc}
\end{figure}

In an out-of-equilibrium system, the correlation functions evolve dynamically. For example, the expectation value of some generic observable, $\cO[\Phi]$, evolves as  
\bea
\av{\cO}_t=\Tr\left[ \rho(t) \cO \right]\,
\ea
with $\rho$ being the density matrix. Given the time evolution of the density matrix, \bea
\rho(t)=U_{t,-\infty} \rho(-\infty) U_{-\infty,t}
\ea
 where $U_{t_f,t_i}=U^\dagger_{t_i,t_f}$ is the time evolution operator between $t_i$ and $t_f$, this expression can be re-written as 
\bea
\av{\cO}_t&=&\Tr\left[  U_{t,-\infty} \rho(-\infty) U_{-\infty,t} \cO\right]=\Tr\left[   \rho(-\infty) U_{-\infty,t} U_{t,\infty}U_{\infty,t}  \cO U_{t,-\infty}\right]\nn
&=&\Tr\left[ \rho(-\infty) U_{-\infty,\infty} U_{\infty,t} \cO U_{t,-\infty}\right] \,.
\ea  
This expectation value has the following path integral representation:
\bea
\av{\cO}_t= \int {\cal D}\Phi \,e^{i \int_{\cal C} L}\, \cO(t)\,,
\ea
where the contour, ${\cal C}$ over which the action is defined is known as the closed-time contour and is shown in Fig.~\ref{fig:cpc}.
 
 More generally, an $n-$point correlation function can be represented by $n$ operator insertions along the contour ${\cal C}$. Regardless of the value of the time coordinate of each operator, the path ordering in the correlation function is determined by the location of the operator along ${\cal C}$. In other words, the lower part of the contour has greater ``contour time" than the upper part. For this reason, in general, for certain time orderings of the operators in the higher point correlation functions, one might need multiple foldings of the contour ${\cal C}$. Here we will restrict ourselves to the class of correlation functions where only two foldings is sufficient. Given the action is given by a contour integral, $I=\int_{\cal C} L$, it is useful to separate the hydrodynamic field $\Phi$ into two components where the time index is evaluated at the upper and lower part of the contour as $\Phi_u$ and $\Phi_d$. Then the contour integral simply becomes 
\be
I[\Phi_u,\Phi_d]=\int_{-\infty}^\infty dt(L[\Phi_u]- L[\Phi_d])
\ee  

We can introduce source fields, $\phi_{u/d}$ that insert an operator on the upper lower part of ${\cal C}$, namely
\be
I[\Phi_u,\Phi_d,\phi_u,\phi_d]=\int_{-\infty}^\infty dt(L[\Phi_u]- L[\Phi_d]+\phi_u\cO_u-\phi_d\cO_d)
\ee 
It is generally more convenient to work with a linear combination of these fields, known as the Keldysh rotation,
\bea
\Phi_r=\frac{1}{2}\left(\Phi_u+\Phi_d\right),\quad \phi_r=\frac{1}{2}\left(\phi_u+\phi_d\right), \quad \Phi_a=\Phi_u-\Phi_d,\quad \phi_a=\phi_u-\phi_d
\ea 
After the Keldysh rotation, the coupling to source term becomes
\be
\phi_u\cO_u-\phi_d\cO_d=\cO_r\phi_a+\cO_a\phi_r,
\ee
therefore an operator $\cO_r$ can be inserted by differentiating the generating functional with respect to $\phi_a$ and vice versa. Furthermore, unitarity of the time evolution operator, $U$, constrains the general form of the action:
\bea
I[\Psi_r,\Psi_a=0]=0,\quad I[\Psi_r,\Psi_a]=-I^*[\Psi_r,-\Psi_a],\quad {\rm Im}S[\Psi_r,\Psi_a]\geq 0
\label{eq:SK_uni}
\ea
where $\Psi$ is an umbrella term that represents both $\Phi$ and $\phi$. All these equations are a direct consequence of the fact that $U$ is a unitary operator, $\Tr \rho=1$, and $0\leq\langle n| \rho |n\rangle\leq1$.

Now we turn to writing down an action for the EFT consistent with the symmetries of the problem. Because the long wavelength modes that remain after the fast modes are integrated out are conserved quantities, the classical equations of motion associated with $I$ correspond to the conservation equations of these quantities. A convenient way to organize the EFT degrees of freedom accordingly is the well known St\"ukelberg trick which we discuss in the next section. 

\subsection{Effective action for nonlinear diffusion}
In sections \ref{sec:stochastic} and \ref{sec:hydrokinetics} we discussed the nonlinear diffusion problem to illustrate how the stochastic approach and deterministic approaches work. In the same spirit, the same diffusion problem was studied in Ref. \cite{Liu:2018kfw} from the EFT perspective which we outline here. The diffusion equation follows from the conservation of the charge current $J^\mu$. The idea behind the St\"ukelberg trick that leads to an action principle is to introduce an external source field, $A_\mu$, that couples to the current in the usual way,
\bea
J^\mu=\frac{\delta I}{\delta A_\mu}\,.
\ea 
What we want is the current conservation, $\del_\mu J^\mu=0$, to be the equation of motion that is obtained by the variation of the action, $I$. This can be achieved by demanding that $I$ depends on the hydrodynamic field $\Phi$ as $I[\Phi,A_\mu]=I[A_\mu+\del_\mu \Phi]$. It is straightforward to observe that the equation of motion for $\Phi$ is simply 
\bea
0=\del_\mu \left(\frac{\delta I}{\delta \del_\mu \Phi}\right)=-\del_\mu \left(\frac{\delta I}{\delta A_\mu}\right)=-\partial_\mu J^\mu
\ea
Note that by construction $I[A_\mu+\del_\mu \Phi]$ is invariant under the gauge transformation $A_\mu\rightarrow A_\mu-\del_\mu \Lambda$, $\Phi \rightarrow \Phi+\Lambda$ which the conservation law is a consequence of. The ``gauge field" here is not dynamical; it is a source term that generates connected correlation functions of the current upon differentiation. As discussed earlier, we actually have \textit{two} source fields, $A_{r,\mu}$ and $A_{a,\mu}$. For  example a two-point function can be calculated as 
\bea
\av{J_{i}^\mu(x_1) J_{j}^\nu(x_2)}= -\left.\frac{\delta^2 {\cal W}[A_{r},A_{a}] }{\delta A_{\bar i \mu}(x_1)\delta A_{\bar j \nu}(x_2)} \right|_{A=0}
\label{eq:JJ_SK}
\ea
where we used the notation $i\in\{r,a\}$, $\bar r=a,\bar a=r$. Here $W$ is the usual generating functional
\bea
e^{{\cal W}[A_{r},A_{a}]}= \log \int {\cal D}\Phi_r {\cal D}\Phi_a \,e^{iI[\Phi_r,\Phi_a,A_r,A_a]}\,.
\ea
Any time ordering of $t_1$ and $t_2$ can be built with the appropriate linear combination of the correlators in Eq.~\eqref{eq:JJ_SK}. In particular, the $rr$, $ra$ and $ar$ components give the symmetric, retarded and advanced correlation functions respectively. The $aa$ component vanishes as a direct consequence of the first equality in Eq.~\eqref{eq:SK_uni} 

The system that describes a normal fluid phase (as opposed to a superfluid) has an extra shift symmetry that further constrains the action \cite{Crossley:2015evo,Glorioso:2017fpd}.  The key idea is that the gauge invariant combination\footnote{The vector $B_\mu$ and its spatial cousin ${\bf B}$ should not be confused with magnetic field despite use of the letter $B$. This review contains no magnetic fields.}
\bea
B_{i\mu}=A_{i\mu}+\del_\mu \Phi_i
\ea
can be viewed as a local source that contains both the external field as well as the hydrodynamic degrees of freedom. In the normal fluid phase this local source has a global (i.e. time independent) diagonal shift symmetry 
\bea
{B}_{0}\rightarrow {B}_{0} ,\quad{{\bf B}}_{u}\rightarrow {{\bf B}}_{u}-{\nabla} \lambda \,,\quad{{\bf B}}_{d}\rightarrow {{\bf B}}_{d}-{\nabla} \lambda .
\ea  
This shift symmetry stems from the phase rotation freedom of each local fluid element and is spontaneously broken in the superfluid phase. We shall focus on the normal phase where the shift symmetry is unbroken. In the $r,a$ basis the shift symmetry becomes
\bea
\label{eq:shift_diff}
{B}_{r0}\rightarrow {B}_{r0} ,\quad{{\bf B}}_{r}\rightarrow {{\bf B}}_{r}-{\nabla} \lambda,\quad B_{a\mu}\rightarrow B_{a\mu} 
\\
{\rm or}\quad \Phi_r\rightarrow \Phi_r-\lambda({\bf x}),\quad \Phi_a\rightarrow \Phi_a
\ea  
Finally, the condition for local thermal equilibrium leads to an additional symmetry which we discuss in the next section.

\subsection{Local Kubo Martin Schwinger (KMS) symmetry}

A system in thermal equilibrium is described by the density matrix given by
\bea
\rho=\frac{e^{-\beta H}}{\Tr (e^{-\beta H})},
\label{eq:rho_thermal}
\ea 
where $H$ is the Hamiltonian and $\beta=1/T$ is the inverse temperature in units where the Boltzmann constant is set to 1. This form of the density matrix leads to a further symmetry, known as the Kubo-Martin-Schwinger (KMS) symmetry which constrains the most general form of the EFT action further, in addition to the constraints imposed by unitarity given in Eq.~\eqref{eq:SK_uni}. 
\begin{figure}[h]
\center
\includegraphics[scale=0.6]{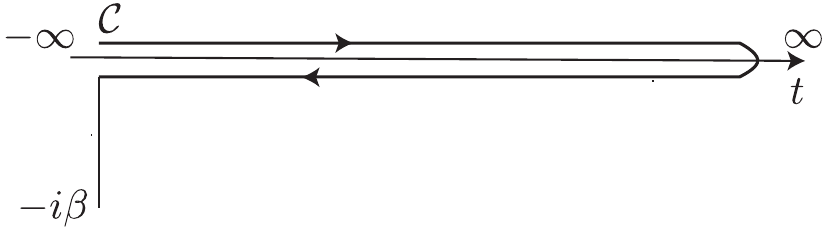}
\caption{The Schwinger-Keldysh contour for systems in thermal equilibrium. The real and imaginary part capture the time evolution and the thermal density matrix respectively. Periodic (anti-periodic) boundary conditions at the end points are assumed for bosonic (fermionic) fields. }
\end{figure}

In order to demonstrate the KMS symmetry we introduce the following notation for the two-point functions\footnote{For brevity we suppress all the indices and arguments expect for time in our notation.}
\bea
G_{<}(t)&:=&\langle\cO(t_2)\cO(t_1)\rangle_{\beta}
=\Tr\left[\rho\cO(t_2)\cO(t_1) \right]
=-\left. \frac{\delta^2{\cal W}[\phi_u,\phi_d]}{\delta \phi_u(t_1)\delta \phi_d(t_2)}\right|_{\phi_u=\phi_d=0}, 
\\
 G_{>}(t)&:=&\langle\cO(t_1)\cO(t_2)\rangle_{\beta}
 =\Tr\left[\rho\cO(t_1)\cO(t_2) \right]
=-\left.\frac{\delta^2{\cal W}[\phi_u,\phi_d]}{\delta \phi_u(t_2)\delta \phi_d(t_1)}\right|_{\phi_u=\phi_d=0} \,.
\ea
where we assumed time translation symmetry and defined $t=t_2-t_1$. The density matrix given in Eq.~\eqref{eq:rho_thermal} has the same form of the time translation operator, albeit with and imaginary time interval equal to the inverse temperature $\beta$. This property relates different two-point functions, 
\bea
G_{>}(t-i\beta)=G_{<}(t)
\label{eq:KMS_det_bal_t}
\ea
 known as the KMS relation. After Fourier transforming into the frequency space, $\omega$, we can see that the KMS relation corresponds to detailed balance,
\bea
G_{>}(\omega)e^{-\beta \omega} =G_{<}(\omega)\,, 
\label{eq:KMS_det_bal_om}
\ea
which ensures thermal equilibrium.  Furthermore, in terms of the retarded, $G_R(t)=i\theta(t)(G_>(t)-G_<(t))$, and symmetric, $G_S(t)=(G_>(t)+G_<(t))/2$, two-point functions, the KMS condition can also be expressed as
\bea
{\rm Im}G_R(\omega)=\coth\left(\frac{\beta\omega}{2}\right)G_S(\omega)
\label{eq:KMS_FDT}
\ea
which is nothing but the fluctuation-dissipation relation where the retarded and symmetric correlators describe dissipation and fluctuations respectively. In fact, in thermal equilibrium, \textit{all} the two point functions, $G_{uu},G_{dd},G_{ud}$,  as well as three point functions can be expressed in terms of the retarded correlator. At the same time, $n\geq4$ point functions can be decomposed in terms of retarded and advanced  components \cite{hou_wang_heinz,wang_heinz}. 

From Eq.~\eqref{eq:KMS_det_bal_t} it follows that the KMS relation involves time reversal. Now suppose the system possesses time reversal symmetry since $<$ and $>$ correlators have opposite time ordering. More generally, let $\Theta$ denote a discrete ${\mathbb Z}_2$ symmetry that involves time-reversal ${\cal T}$. Depending on the system in question, it could be just ${\cal T}$, or include some combination of parity and charge conjugation \cite{Crossley:2015evo,Glorioso:2017fpd}. The KMS condition for such a system defines a symmetry where the generating functional remains invariant under the transformation
\bea
\phi_u(t)\rightarrow \Theta\phi_u(t-i\theta),\quad \phi_d(t)\rightarrow \Theta\phi_d(t+i(\beta-{\varphi}))
\label{eq:KMS_ud}
\ea 
 for any $\varphi\in[0,\beta]$. The existence of the arbitrary parameter is due to the time-translation symmetry. It is straightforward to show that after performing the Keldysh rotation, in the classical limit, $\hbar\rightarrow 0$ the KMS transformation simply becomes
  \bea
\phi_r\rightarrow \Theta\phi_r,\quad \phi_a\rightarrow \Theta\phi_a+i\Theta\beta\partial_t\phi_r
\ea
At this stage the KMS symmetry is expressed via the correlation functions and hence is a property of the generating functional ${\cal W}$. It can be shown that ${\cal W}$ satisfies the symmetry if we require that the full action satisfies
\bea
I[\Psi_r,\Psi_a]=I[\Theta\Psi_r,\Theta\Psi_a+i\Theta\beta\partial_t\Psi_r]
\ea
where both the dynamical fields, $\Phi$ and the sources transform under the KMS transformation \cite{Crossley:2015evo}.   

In particular, for the diffusion example we discussed above, the KMS symmetry implies that the action is invariant under the transformation:
\bea
B_{r\mu}(x) \rightarrow  B_{r\mu}(-x),\quad B_{a\mu}(x)\rightarrow  B_{a\mu}(-x)-i\beta  \partial_t B_{r\mu}(-x)\,.
\label{eq:KMS_diff}
\ea
We note that $\Theta B(x)=B(-x)$ and $\Theta \del_\mu=-\del_\mu$. Following the usual philosophy of EFT, we can write down the action consistent with the underlying symmetries, Eqs. \eqref{eq:SK_uni}, \eqref{eq:shift_diff} and \eqref{eq:KMS_diff} as an expansion in fields and derivatives. It is useful to observe that the combination, $B_{a\mu}(B_{a\mu}+i\beta\del_tB_{r\mu})$ ($\mu$ not summed over) is invariant under the KMS transformation given in Eq.~\eqref{eq:KMS_diff}.  In addition, requiring translational and rotational symmetry, we then conclude that the most general form of the lagrangian up to quadratic order in $B_a$ field and linear order in derivatives contains two terms: $B_{a0}$ and $i {{\bf B}_{a}}\cdot \left({\bf B}_{a}+i\beta \del_t{\bf B}_{r}\right)$. The coefficients of these terms can be arbitrary functions of $B_{r0}$, which is identified as the local chemical potential $\mu$. At the same time, it is  convenient to use $\alpha=\beta B_{r0}=\beta \mu$ instead as the dynamic variable. Consequently, the most general form of the action to leading order in the EFT is given by
\bea
{\cal L}[B_r,B_a]=n(\alpha)B_{a0}+i\lambda(\alpha){{\bf B}_{a}}\cdot \left({\bf B}_{a}+i\beta \del_t{\bf B}_{r}\right)\,.
\label{eq:eft_diff}
\ea
where at this point $n$ and $\lambda$ are generic functionals of $\alpha$. The unitarity constraint implies that $\lambda\geq 0$. Let us now discuss the underlying physics of this system. If we turn off the sources $A_a=A_r=0$, by using $B_{r\mu}=\del_\mu \Phi_r$, we can write $\del_t {\bf B}_{r}=\nabla \mu$ and obtain 
\bea
{\cal L}[\alpha,\Phi_a]&=&n(\alpha)\del_t \Phi_a +i\lambda(\alpha){\nabla \Phi_{a}}\cdot \left(\nabla \Phi_{a}+i\nabla\alpha \right)\\
&=&-\Phi_a \left(\del_t n(\alpha)-\nabla \cdot \left(\lambda(\alpha)\nabla \alpha\right) \right)  +i \Phi_{a}\nabla \cdot  \left(\lambda(\alpha)\nabla  \right) \Phi_a\\\
&=&-\Phi_a \left(\del_t n(\alpha)+\nabla \cdot {\bf J}_D \right)  +i \Phi_{a}\nabla \cdot  \left(\lambda(\alpha)\nabla  \right) \Phi_a\
\label{eq:EFT_diff}
\ea
where in the second line we integrated by parts. We see that the equation of of motion, to leading order in $\Phi_a$, is the classical conservation law (no noise), 
\bea
\del_t n(\alpha)+\nabla \cdot {\bf J}_D=0
\ea
where notably the dissipative, constitutive current,  
\bea
 {\bf J}_D=-\lambda(\alpha)\nabla \alpha,
\ea
appears as a consequence of the EFT, and not as an empirical law as it was introduced in Eq.~\eqref{eq:constitutive_J}. In this form, the constitutive relation describes non-linear diffusion where the density, $n$ and chemical potential $\alpha=\mu/T$ are related by some general equation of state $n(\alpha)$. From the EFT perspective the susceptibilities that arise when we expand $n$ in powers of $\alpha$ can be viewed as some unspecified coupling constants. Keeping only the linear term, $n= T\chi \alpha$, with $\chi$ being the susceptibility, reproduces ordinary diffusion. The diffusion constant is simply 
\bea
D=\frac{\lambda}{\chi T}
\ea 

 The second term in Eq. \ref{eq:EFT_diff} that is quadratic in $\Phi_a$ corresponds to stochastic noise. To see this more clearly let us first introduce a vector-like Hubbard-Stratonovich field ${\bf I}(x)$ and integrate the $\Phi_a$ field:
  \bea
 Z&=&\int {\cal D}\Phi_r {\cal D}\Phi_a \,
\exp\left[i \int d^4x\, \left( -\Phi_a \left(\del_t n(\alpha)+\nabla \cdot {\bf J}_D \right) 
+i  \lambda(\alpha)\nabla \Phi_a\cdot\nabla \Phi_a \right) \right] \\
 &
=& 
\int {\cal D}\Phi_r {\cal D}\Phi_a  {\cal D}{\bf I} \,
 \exp\left[i\int d^4x\, \left(-\Phi_a \left(\del_t n(\alpha)+\nabla \cdot {\bf J}_D \right) +   {\bf I} \cdot\nabla\Phi_a +i  \frac{|{\bf I}|^2}{4\lambda}  \right)\right]\\
  &
=& 
\int {\cal D}\Phi_r {\cal D}\Phi_a  {\cal D}{\bf I} \,
 \exp\left[i\int d^4x\, \left(-\Phi_a \left(\del_t n(\alpha)+\nabla \cdot ({\bf J}_D+{\bf I}) \right)  +i  \vec \frac{|{\bf I}|^2}{4\lambda}  \right)\right]
  \ea
 where in the last line we integrated by parts the second term. Now, the $\Phi_a$ integration is trivial and leads to a (functional) delta function
 \bea
 Z&=&\int {\cal D}\Phi_r {\cal D}{\bf I} \, \delta\left[\del_t n+\nabla \cdot\left( {\bf J}_D+{\bf I} \right)\right]\,\exp\left[- \int d^4 x\frac{|{\bf I}|^2}{4\lambda} \right]
 \label{eq:MSR}
 \ea
 This is nothing but the path integral representation of the Langevin equation described in Section \ref{sec:stochastic} with ${\bf I}$ being the noise term which satisfies
 \bea
 \av{I^i(\vx)I^j(\vy)}= \int  {\cal D}{\bf I}  I^i(\vx)I^j(\vy) \exp\left[- \int d^4 x\frac{|{\bf I}|^2}{4\lambda} \right] =2\lambda \delta^{(4)}(x-y)\delta^{ij}
 \ea 
 in accordance with the fluctuation dissipation theorem. It is also known as the Martin-Siggia-Rose form \cite{Martin:1973zz}. But rather than being empirically postulated, it arises from the general EFT considerations. 
 
 One technical point is that in the Schwinger-Keldysh formalism the partition function represents the trace of the density matrix which is unity. This translates into the normalization condition \cite{Kamenev_2011}
 \bea
 Z[{\bf I}]:=\int {\cal D}\Phi_r j[\Phi_r ]\, \delta\left[\del_t n+\nabla \cdot\left( {\bf J}_D+{\bf I} \right)\right]=1
  \ea 
 where $j[\Phi_r ]$ is the jacobian of the transformation $\Phi_r\rightarrow  \del_t n+\nabla \cdot\left( {\bf J}_D+{\bf I} \right)$. The different ways of regularizing the operator that leads to the jacobian correspond to different discretization prescriptions discussed in Section \ref{sec:stochastic}. The It$\bar{\rm o}$ prescription corresponds to the discretization where the operator is upper triangular and hence $j[\Phi_r ]=1$. In general, it is possible to exponentiate the jacobian by using fermionic ghosts. This operation results in a BRST invariance in where the theory is invariant under a set of supersymmetries that mix the  fermionic ghost fields and the bosonic $\Phi_r$, $\Phi_a$ fields. The transformation that leads to the jacobian is known as the Nicolai map in supersymmetric field theory \cite{NICOLAI1980419,NICOLAI1980341} and its implications for stochastic systems have been discussed in \cite{Ezawa:1984bm}. More recently, in the context of hydrodynamic EFTs the relation between BRST symmetry and unitarity have been discussed in detail in \cite{Crossley:2015evo,Haehl:2016pec,Haehl:2016uah,Jensen:2017kzi}. In a later paper, it was argued that there is a particular regularization scheme where the fermionic ghosts decouple and BRST symmetry is not needed \cite{Gao:2018bxz}.
 
\subsection{Effective action for hydrodynamics}
We now turn to  full hydrodynamics where the relevant conserved quantity is the energy-momentum tensor. In the EFT framework, just like the conserved current is given by the variation of the action with respect to a gauge field in the diffusion example, the conserved energy-momentum tensor is given by the variation of the action with respect to a background metric
\bea
T^{\mu\nu}=\frac{\delta I}{\delta g_{\mu\nu}}
\ea 
In the closed time path formalism we have two copies of the metric, $g_{u,\mu\nu}$ and  $g_{d,\mu\nu}$ which source the energy-momentum tensor on the upper and lower parts of the Schwinger-Keldysh contour. In this case, the symmetry associated with the conservation law which leaves the action invariant is space-time diffeomorphisms. The EFT is organized around this symmetry in combination with the KMS symmetry. In order to promote the generators of diffeomorphisms into dynamical fields, a formulation of fluid dynamics have been introduced in \cite{Crossley:2015evo,Glorioso:2017fpd} which we summarize below.

 In this formulation fluid space-time is parameterized by an auxiliary space-time variables $\sigma^A=( \sigma^0,\sigma^i)$. The spatial components, $\sigma^i$ with $i\in \{1,2,3\}$, parameterize the position of the fluid element, and $\sigma^0$ defines an internal clock carried by each fluid element. For a fluid, in addition to space-time diffeomorphisms, there are also two extra diffeomorphism invariances, namely a time independent one $\sigma^i\rightarrow \sigma^{i\prime}(\sigma^i)$, and a re-parameterization of the internal time: $\sigma^0\rightarrow \sigma^{0\prime}(\sigma^0,\sigma^i)$. These essentially ensure that each fluid element is free to move independently, as opposed to a solid \cite{Liu:2018kfw}. 
 
 The (two copies) of physical space-time, $X^\mu_u(\sigma^A)$, $X^\mu_d(\sigma^A)$, are dynamical fields and they define a relativistic version of the Lagrangian description of a fluid. The hydrodynamic fields are derived from them. For example the fluid velocity is given by the time-like vector 
 \bea
 u_i^\mu = \frac{1}{|  \del X_i/\del \sigma^0|} \frac{\del X_i^\mu}{\del \sigma^0}
  \ea
The Keldysh rotated combinations correspond to the physical space-time, $X_r^\mu:=x^\mu$ and a vector-valued noise field $X_a^\mu$.  The physical space-time metric in the Keldysh basis is similarly defined as $g_{r\mu\nu}=g_{\mu\nu}$.  The connection to the space-time formulation of the fluid can be achieved by inverting the relation and expressing $\sigma(x)$ as a dynamical variable. 

 In a dynamical fluid, the energy density hence the temperature varies over space-time. In the EFT formulation it is useful to define the local inverse temperature as one of the independent fields, $\beta(\sigma(x))=1/T(\sigma(x))$.  Energy density, pressure, entropy, are determined by the equation of state as functions of temperature.  It is convenient to merge the inverse temperature and fluid velocity into an unconstraint vector
\bea
\beta^\mu(x)=\beta(x) u^\mu(x)
\ea
 dubbed as the thermal vector. In terms of the space-time fields it can be defined as 
 \bea
  \beta^\mu(x)=\frac{1}{T_0}\left.\frac{\del x^\mu(\sigma)}{\del \sigma^0} \right|_{\sigma=\sigma(x)}
 \ea

Putting all these symmetry considerations together leads to the following basic building blocks for the EFT:
\bea
\beta(x), \,u^\mu(x), \, \bD^{\mu\nu}, \, G_{r\mu\nu}:=g_{\mu\nu}, \, G_{a\mu\nu}:=g_{a\mu\nu}+{\cal L}_{X_a} g_{\mu\nu},
\ea
where ${\cal L}_v$ denotes the Lie derivative along the vector $v$ and $g_{a\mu\nu}$ is the Keldysh rotated metric which correspond to the source field for $T^{\mu\nu}_a$. The final symmetry we need to impose is the KMS symmetry, which ensures thermal equilibrium. For the case of a fluid flow which involve flow of energy, the thermal equilibrium condition is \textit{locally} imposed. Therefore the resulting KMS symmetry is also expressed such that the temperature each fluid cell is determined locally. Following Ref. \cite{Liu:2018kfw}, this local version of the KMS transformation can be written succinctly as
\bea
u^\mu(x)\rightarrow \Theta u^\mu(x),\quad \beta(x)\rightarrow \Theta \beta(x),\quad G_{a\mu\nu}(x)\rightarrow \Theta G_{a\mu\nu}(x)+ i\Theta \Lb G_{r\mu\nu}(x)
\ea
Similar to Eq.~\eqref{eq:eft_diff} for nonlinear diffusion, the effective action up to first order in gradients can be written as a sum of two terms, each of which are separately invariant under KMS transformations 
\bea
{\cal L}&=&\frac12(\eps u^\mu u^\nu+p \bD^{\mu\nu})G_{a\mu\nu}+i\frac{T}{4}\left[ \eta(\bD^{\mu\kappa}\bD^{\nu\lambda}+\bD^{\mu\lambda}\bD^{\nu\kappa})+\left(\zeta-\frac{2}{3} \eta\right) \bD^{\mu\nu}\bD^{\lambda\kappa}\right] G_{a\mu\nu}\left(G_{a\lambda\kappa}+i\Lb G_{r\lambda\kappa}\right)\nn
\label{eq:eft_hydro}
\ea
The physical interpretation of this expression is straightforward: the first term encodes the ideal contribution to the energy momentum tensor, and the second term encodes both the fluctuations and the dissipation. Notice that the coefficient that appears in square brackets is precisely the term that appears in the noise correlator in the stochastic approach given in Eq.~\eqref{eq:gaussian_noise_Pi}. As an illustration of how this works, let us consider Minkowski space, $g_{\mu\nu}=$diag$(-1,1,1,1)$. Turning off the source $g_{a\mu\nu}=0$ and expressing the Lie derivatives explicitly we get
\bea
G_{a\mu\nu}&=&g_{a\mu\nu}+{\cal L}_{X_a} g_{\mu\nu}=\nabla_\mu X_{a\nu} + \nabla_\nu X_{a\mu}\\
\Lb G_{r\mu\nu}&=&\Lb g_{\mu\nu}=\nabla_\mu \beta_\nu+ \nabla_\nu \beta_\mu 
\ea
When contracted with the projection operators, the terms that are proportional to the gradient of temperature vanish in the second line and this term reproduces $\Pi^{\mu\nu} G_{a\mu\nu}$ where $\Pi^{\mu\nu}$ is the dissipative part of the energy momentum tensor defined in Eq.~\eqref{eq:dissipative_Pi}. Upon integrating by parts the $\nabla_\mu X_{a\nu}$ terms we finally obtain obtain
\bea
{\cal L}&=&- X_{a\nu}\nabla_\mu T^{\mu\nu}+i\frac{T}{2} (\nabla_\mu X_{a\nu}) \left[ \eta(\bD^{\mu\kappa}\bD^{\nu\lambda}+\bD^{\mu\lambda}\bD^{\nu\kappa})+\left(\zeta-\frac{2}{3} \eta\right) \bD^{\mu\nu}\bD^{\lambda\kappa}\right]  (\nabla_\lambda X_{a\kappa})
\ea
where $T^{\mu\nu}$ contains both the ideal and the dissipative parts. Just as in the nonlinear diffusion problem, this Lagrangian is nothing but the MSR form of the stochastic relativistic hydrodynamics described in Section \ref{sec:stochastic}. This MSR form of the effective action, along with a prescription for discretization has been introduced in \cite{Kovtun:2014hpa}.

\subsection{KMS blocks and new transport coefficients}
\label{sec:KMS_blocks}

So far the EFT formalism allowed us to reproduce stochastic hydrodynamics, discussed in Section \ref{sec:stochastic}, albeit from an organizing principle based on general symmetry considerations. This is not very surprising, since both formalisms are constructed to obey the fluctuation-dissipation theorem which fixes the magnitude of the noise in terms of the dissipative transport coefficients. More precisely, as discussed earlier, fluctuation-dissipation theorem completely determines the two and three point functions in terms of the retarded correlator; and classical hydrodynamics essentially determines the form of the retarded correlators. The fluctuation effects, such as long-time tails can be calculated via loops of these basics retarded correlators.

 At the same time, if one is interested in studying non-retarded higher point functions to obtain some information about the dynamics of the system, the fluctuation-dissipation theorem is not sufficient. In this case, more information beyond the transport coefficients of classical hydrodynamics is necessary.  The EFT formalism is tailored to systematically study such stochastic contributions which are not a part of classical hydrodynamics. This was done in the notable Ref. \cite{Jain:2020zhu} by classifying the higher order (both in gradients and in the number of $a$ fields) terms in the effective action for hydrodynamics. 
 
 The main idea pursued in Ref. \cite{Jain:2020zhu} stems from the observation that in Eq.~\eqref{eq:eft_diff} and Eq.~\eqref{eq:eft_hydro}, each term is invariant under the dynamical KMS transformation. In particular, the second terms in each equation encode both dissipation and fluctuations with a single coefficient as a result of the fluctuation-dissipation theorem. The coefficients of the dissipative parts are, by definition, given by transport coefficients of classical hydrodynamics. As a consequence of the KMS symmetry, they then fix the form of the fluctuation terms as well. It is then possible to organize the higher order terms in the effective action in similar groups, each of which are separately invariant under the KMS transformation. These groups are named as ``KMS blocks". The coefficients of each KMS block is a new stochastic coefficient, independent of the transport coefficients given by classical hydrodynamics. This is because the blocks do not mix via KMS which enforces fluctuation dissipation theorem. 
 
 For the nonlinear diffusion, there are two KMS blocks to next-to-leading order (hence the subscript 2) which are given by  
\bea
{\cal L}^{\rm diff}_2=i\vartheta_1B_{ai} B_{aj} \left(\del_tB^i_{r} \del_t B^j_{r}-\delta^{ij}\del_t{\bf B}_{r} \cdot \del_t{\bf B}_{r} \right)+iT^2\vartheta_2\left({\bf B}_a\cdot  {\bf B}_a\right) \left({\bf B}_a+i \beta \del_t {\bf B}_{r}\right) \cdot \left({\bf B}_a+i \beta \del_t {\bf B}_{r}\right)\,.
\ea
Note that each term has a separate coefficient $\vartheta_1$ and $\vartheta_2$ which are in general functions of $\alpha$. These are not determined by the equation of state, and they have to be determined from some additional short distance information from the underlying microscopic theory. For hydrodynamics, the KMS blocks are of similar form: 
\bea
{\cal L}^{\rm hydro}_2&=&
i\vartheta_3\left(\bD^{\mu\mu^\prime}\bD^{\nu\nu^\prime}\bD^{\lambda\lambda^\prime} \bD^{\kappa\kappa^\prime} -\bD^{\mu\lambda}\bD^{\nu\kappa}\bD^{\mu^\prime\lambda^\prime} \bD^{\nu^\prime\kappa^\prime}\right)
\G_{a \mu\nu} \G_{a\lambda\kappa} \Lb G_{r \mu^\prime\nu^\prime} \Lb G_{r\lambda^\prime\kappa^\prime}
\nn
&&+iT^2\vartheta_3 \bD^{\mu\nu}\bD^{\mu^\prime\nu^\prime}\bD^{\lambda^\prime\kappa^\prime}G_{a\mu\nu}\G_{a\mu^\prime\nu^\prime}\left(G_{a\lambda\kappa}+i\Lb G_{r\lambda\kappa}\right) 
\left(G_{a\lambda^\prime\kappa^\prime}+i\Lb G_{r\lambda^\prime\kappa^\prime}\right)
\ea
Note that in either case they contain either four $a$ fields, encoding non-Gaussian noise, or higher order gradients compared to the leading order terms in the EFT given in Eqs. \ref{eq:eft_diff} and \ref{eq:eft_hydro}. The general form of the higher order KMS blocks are worked out in \cite{Jain:2020zhu}. 

The effects of these new transport coefficients appear at two-loops in the retarded two point correlator and proportional to $\omega k^2$ (whereas the correction due to the long-time tails goes as $\omega^{1/2}k^2$). For the retarded three point function their contribution is non-analytic, similar to the long-time tails but as we discussed,  unlike the long-time tails, the coefficient is not fixed by the transport coefficients that appear in the constitutive relations. Although they are higher order in gradients, these new terms are still within the hydrodynamic limit, and they might potentially lead to observable effects.  
 
\subsection{Equal time correlation functions and Schwinger-Dyson equations}

Having an effective action for stochastic hydrodynamics means that it is possible to obtain the evolution equations for fluctuations starting from the action. The classical equations of motion, to leading order in $\Phi_a$, correspond to the evolution equations for the one-point functions, namely the ordinary hydrodynamic equations without fluctuations. Just as in quantum field theory, the evolution equations of the two and higher point functions follow from the Schwinger-Dyson equations of the effective field theory. This approach to the hydro-kinetic equations has been taken in Ref. \cite{Sogabe:2021svv} in the context of nonlinear diffusion. We outline the main idea below. 
 
Let us consider the average charge density, 
\bea
\av{ n(x)}=\int {\cal D}n {\cal D}\Phi_a   n(x) e^{i I[n,\Phi_a]}\,.
\ea
where the action $I[n,\Phi_a]$ is given in Eq.~\eqref{eq:EFT_diff}. Here we switched our primary variable and explicitly denoted the hydrodynamic field $n$ in the path integral measure instead of $\Phi_r$. Following the standard textbook argument, the path integral is invariant under local reparameterizations of the noise field  $\Phi_a$, 
\bea
\Phi_a(x)\rightarrow \Phi_a(x)+\delta \Phi_a(x)
\ea
which leads to the relation
\bea
0=\langle \frac{\delta I}{\delta \Phi_a}(x_1)  n(x_2)\rangle=
 \av{ \left( \del_{t_1} n(x_1)+\nabla \cdot {\bf J}_D(x_1) +i\nabla \cdot  (\lambda\nabla ) \Phi_a(x_1) \right)  n(x_2) }\,.
\ea
Reorganizing the terms and subtracting the average yields 
\bea
\av{ \del_{t_1} \delta n(x_1)  \delta n(x_2) }=
 -\av{ F(x_1) \delta n(x_2)} +i \av{Q(x_1)\Phi_a(x_1)  \delta n(x_2) }\,.
 \label{eq:SD_general}
\ea
where $\delta n=n-\av{n}$ denoted the deviation from the average as usual and the drift and noise terms are given similarly to Langevin equation Eq.~\eqref{eq:dtn} as 
\bea
 F&=&-\nabla\cdot \left(\lambda[n]\nabla \alpha[n]-\av{\lambda[n]\nabla \alpha[n]}\right)=-\gamma \nabla^2 \delta n+ {\cal O}(\delta n^2,\eps_q^3) \nn
 Q&=&-\nabla \cdot  (\lambda[n]\nabla ) =-\lambda \nabla^2+{\cal O}(\delta n,\eps_q^3)\,
\label{eq:SD_FQ}
\ea
 Notice that because $\lambda$ and $\alpha$, and hence $ F$ and $Q$, depend nonlinearly on $\delta n$, three and higher point functions appear on the right hand side of Eq.~\eqref{eq:SD_general} in the same way as discussed Sec. \ref{sec:nongaussian}. They are higher order in $\eps$ in the hydrodynamic limit.  We focus on the leading behavior of $F$ and $Q$ in the $\eps$ expansion which is given in the right hand side in Eq.~\eqref{eq:SD_FQ}. Note that in the notation used in Eq.~\eqref{eq:SD_FQ}, the transport coefficients $\gamma$ and $\lambda$ on the left hand side are nonlinear functionals of the fluctuating density, whereas on the right hand side they are functionals of the average density, i.e. $\gamma:=\gamma[\av{n}]$,  $\lambda:=\lambda[\av{n}]$. Because they do not depend on the fluctuating field, their spatial variation is controlled solely by the evolution of the background (i.e. the average field) and not the fluctuations. Consequently, terms that involve the gradients of $\gamma$ and $\lambda$ are suppressed compared to the terms that involve the gradients of the fluctuating fields. In what follows, we will neglect the terms that involve the gradient of background terms such as $\nabla \lambda$ which are of order $\eps_q^2$, whereas the typical gradient of $\delta n$ is of order $\eps_q$.  

The contribution of noise contribution in Eq.~\eqref{eq:SD_general} to leading order in $\eps$ follows from the causal structure of Schwinger-Keldysh formalism where \cite{Liu:2018kfw}
\bea
\av{\Phi_a(x_1)\delta n(x_2)}=-i\theta(t_2-t_1)[\Phi_a(x_1),\delta n(x_2)]
\ea
Physically this can be understood as the noise at any given time only affects the fields in the future. Finally, in order to calculate the equal time correlators, we need to take the limit $t_1=t_2$ in Eq.~\eqref{eq:SD_general}. The equal time limit can be taken in a symmetric way by considering $(\del_{t_1}+\del_{t_2}) \av{\delta n(x_1)\delta n(x_2)}$ such that $t=(t_1+t_2)/2$ \cite{Sogabe:2021svv}. This way, the result is independent of whether the equal time limit is taken from above or below, i.e. $t_1-t_2\rightarrow 0^+$ or $0^-$. Furthermore, from the action given in Eq.~\eqref{eq:EFT_diff}  one can deduce that $\Phi_a$ and $n$ are canonical conjugate fields and therefore satisfy the equal time correlation
\bea
[\Phi_a(t, \vx_1),n(t, \vx_2)]=-i\delta^{3}(\vx_1-\vx_2) 
\ea
Putting everything together and taking the Wigner transform with respect to the fluctuation coordinate $\vx_1-\vx_2$, reproduces the two point function given in Eq.~\eqref{eq:W_234-b} obtained from the deterministic hydro-kinetic formalism.

The evolution equations for the higher point functions can be obtained in a way similar to the two point functions by considering 
\bea
0=\av{\frac{\delta I}{\delta \Phi_a}(x_1)  n(x_2) \dots n(x_n)} 
\ea
which leads to
\bea
  \av{ \del_{t_1} \delta n(x_1)  \delta n(x_2) \dots \delta n(x_n) }=  -\av{ F(x_1)   \delta n(x_2) \dots \delta n(x_n)} +i \av{Q(x_1)\Phi_a(x_1)   \delta n(x_2) \dots \delta n(x_n) }.
  \label{eq:SD_general_n}
\ea
We similarly expand the drift and noise terms both in fluctuations (controlled by $\eps$) and gradients (controlled by $\eps_q$). For the gradient expand, it suffices to expand to leading order in $\eps_q$. However, for fluctuations we need to expand  $F$ and $Q$ beyond linear order in $\delta n$, unlike the two-point function. This is because terms like $\delta n(x_1)^2$ that comes from expanding $F$ or $Q$ beyond linear order, lead to coincident, loop corrections in the two-point function equation. In the equations for higher point functions, however, they can be contracted with terms such as $\delta n(x_2)\delta n(x_3)$. The contribution from such disconnected combinations of lower point functions are in fact the same order as connected higher point functions as seen in Eq.~\eqref{eq:Gn_epsilon}. For example, both $G_2\times G_2$ and $G_3$ are of order $\epsilon^2$. Therefore, in order to take into account all terms in leading order in $\epsilon$, $F$ and $Q$ have to be expanded to order $\delta n^{n-1}$, and $\delta n^{n-2}$ respectively. 

Similarly to the two-point function, we take taking the symmetric equal-time limit, i.e.  $(\del_{t_1}+\dots +\del_{t_n})  \av{ \delta n(x_1)  \delta n(x_2) \dots \delta n(x_n) }$ with $t=(t_1+\dots+ t_n)/n$. The noise terms that involve multiple $\delta n$s and a single $\Phi_a$, carry the same causal structure as the two-point function \cite{Gao:2018bxz}. After these terms are evaluated and the multi-point Wigner transform is performed, the Schwinger-Dyson equations for the three and four point functions match exactly with the evolution equations given in Eq.~\eqref{eq:W_234-b}. As one would expect, all the drift terms (denoted by the terms with the semicircle vertices in Fig.~\ref{fig:diagram}) follow from the first term in the right-hand side of Eq.~\eqref{eq:SD_general_n} and the noise terms (those with triangle vertices) follow from the second term.

Finally, in Ref. \cite{Sogabe:2021svv} the effects of the new noise terms given in terms of the KMS blocks as reviewed in Sec. \ref{sec:KMS_blocks} are also explored. As expected, the contribution of these terms to the evolution equations for any $n-$point function is higher order in the gradient expansion. For consistency, these effects should be taken into account if one wishes to go beyond the leading order in the gradient expansion and consider higher order gradient terms in the constitutive relations. At the same time, as we have seen earlier, the effect of the one-loop (long-time tails) terms are more dominant that the second order terms in the gradient expansion and they have to be included before one goes beyond the leading order in gradients. Of course, the effects of these new noise terms can be more visible in higher order correlators with more complicated time orderings as opposed to equal-time correlators \cite{Jain:2020zhu}.

%% file: conclusions.tex
Despite its old age, relativistic hydrodynamics is still a rapidly developing field. In this review we presented a bird's eye view of the recent progresses in the study of thermal fluctuations in relativistic hydrodynamics from three different but complimentary perspectives. All three of these frameworks have particular advantages and challenges of their own, but they all have one thing in common which is that there are many open questions.

Stochastic hydrodynamic has now reached a stage where the state-of-the-art is 3+1 dimensional simulations \cite{SINGH2019319}. The novel use of the Metropolis algorithm opens a promising avenue for overcoming some of the challenges of the Langevin method, such as multiplicative noise and dealing with short distance singularities. This is a relatively new direction which we expect rapid developments in the near future. In particular, while there are Metropolis simulations of fluid-like systems such as nonlinear diffusion \cite{Basar:2024qxd} and model H \cite{Chattopadhyay:2024jlh} , a full simulation of 3+1 dimensional relativistic hydrodynamics is yet to be carried on. An exciting feature of the Metropolis framework we wish to highlight is that it does not contain dissipative terms as an input; the inclusion of fluctuations automatically generates dissipation. This could, in addition to taking into account thermal fluctuations in the first place, mitigate or even practically eliminate issues with causality and stability inherent in viscous relativistic hydrodynamics.   

The more recent framework of hydro-kinetics also has seen progress over the last few years. The equations for two-point functions with a generic background is now complete \cite{An:2019fdc}. The non-gaussian fluctuations that include three and higher point functions is yet to be completed. The general formalism \cite{An:2020vri} and the equations for the slowest mode \cite{An:2022jgc} have been derived but they lack the inclusion of velocity fluctuations at the moment. In the presence of Lorentz symmetry, this is a challenging task. Given that the short distance singularities and renormalization can be taken care of analytically, and any type fluid background can be included as a source term (if one neglects the long-time tail corrections), these equations can easily be used to make quantitative predictions for the observable signals of fluctuations. This is a crucial ingredient of the search for the QCD critical point via heavy ion collisions.

The effective field theory formulation of hydrodynamics provides a first principles approach based on symmetries and action principle. In addition, it also has led to remarkable results such as the existence of new transport coefficients which have no counterpart in classical hydrodynamics organized by the KMS symmetry \cite{Jain:2020zhu}. The effects of these ``non-classical" sectors have not yet been fully explored. Likewise, the relation between the hydro-kinetic approach and effective field theory has not been fully understood yet either. The first step in this direction was taken in Ref. \cite{Sogabe:2021svv} but the fluctuations of velocity and, in general, how the confluent formalism that describes the fluctuations in a Lorentz covariant way, appears from the effective theory and how does it manifest itself in the Schwinger-Dyson equations is has yet to be understood. 

As in the case of any rapidly evolving and vibrant research field, this review only offers a snapshot of the current lay of the land of the field. There could be other areas, such as chaotic or turbulent systems, cold atoms or dense systems like neutron stars where the study of hydrodynamic fluctuations can potentially have an impact in the future.